\documentclass[aps,prd,twocolumn,superscriptaddress]{revtex4}
\usepackage{epsfig,epsf}
\usepackage{amsmath}
\usepackage{amsthm}
\usepackage{amsfonts}
\usepackage{amssymb}
\usepackage{dsfont}
\usepackage{multirow}
\usepackage{appendix}
\usepackage{slashed}
\usepackage[active]{srcltx}
\usepackage{psfrag}

\begin{document}

\title{Axial-vector and pseudoscalar tetraquarks $[ud][\overline{c}\overline{%
s}]$}
\date{\today}
\author{H.~Sundu}
\affiliation{Department of Physics, Kocaeli University, 41380 Izmit, Turkey}
\affiliation{Department of Physics Engineering, Istanbul Medeniyet University, 34700
Istanbul, Turkiye}
\author{S.~S.~Agaev}
\affiliation{Institute for Physical Problems, Baku State University, Az--1148 Baku,
Azerbaijan}
\author{K.~Azizi}
\affiliation{Department of Physics, University of Tehran, North Karegar Avenue, Tehran
14395-547, Iran}
\affiliation{Department of Physics, Do\v{g}u\c{s} University, Acibadem-Kadik\"{o}y, 34722
Istanbul, Turkey}

\begin{abstract}
Spectroscopic parameters and widths of the fully open-flavor axial-vector
and pseudoscalar tetraquarks $X_{\mathrm{AV}}$ and $X_{\mathrm{PS}}$ with
content $[ud][\overline{c}\overline{s}]$ are calculated by means of the QCD
sum rule methods. Masses and current couplings of $X_{\mathrm{AV}}$ and $X_{%
\mathrm{PS}}$ are found using two-point sum rule computations performed by
taking into account various vacuum condensates up to dimension $10$. The
full width of the axial-vector state $X_{\mathrm{AV}}$ is evaluated by
including into analysis $S$-wave decay modes $X_{\mathrm{AV}}\to D^{\ast
}(2010)^{-}K^{+}$, $\overline{D}^{\ast }(2007)^{0}K^{0}$, $D^{-}K^{\ast
}(892)^{+}$, and $\overline{D}^{0}K^{\ast }(892)^{0}$. In the case of $X_{%
\mathrm{PS}}$, we consider $S$-wave decay $X_{\mathrm{PS}}\to \overline{D}%
_{0}^{\ast }(2300)^{0}K^{0}$, and $P$-wave processes $X_{\mathrm{PS}}\to
D^{-}K^{\ast}(892)^{+}$ and $X_{\mathrm{PS}}\to \overline{D}%
^{0}K^{\ast}(892)^{0}$. To determine partial widths of these decay modes, we
employ the QCD light-cone sum rule method and soft-meson approximation,
which are necessary to estimate strong couplings at tetraquark-meson-meson
vertices $X_{\mathrm{AV}}D^{-}D^{\ast }(2010)^{-}K^{+} $, etc. Our
predictions for the mass $m_{\mathrm{AV}}=(2800 \pm 75)~\mathrm{MeV}$ and
width $\Gamma_{\mathrm{AV}}=(58 \pm 10)~\mathrm{MeV}$ of the tetraquark $X_{%
\mathrm{AV}}$, as well as results $m_{\mathrm{PS}}=(3000 \pm 60)~\mathrm{MeV}
$ and $\Gamma_{\mathrm{PS}}=(65 \pm 12)~\mathrm{MeV}$ for the same
parameters of $X_{\mathrm{PS}}$ may be useful in future experimental studies
of multiquark hadrons.
\end{abstract}

\maketitle

%%%%%%%%%%%%%%%%%%%%%%%%%%%%%%%%%%%%%%%%%%%%%%%%%%%%%%

\section{Introduction}

\label{sec:Int} %%%%%%%%%%%%%%%%%%%%%%%%%%%%%%%%%%%%%%%%%%%%%%%%%%%%%%%
Recent LHCb information on new structures $X_{0}(2900)$ and $X_{1}(2900)$
observed in the invariant $D^{-}K^{+}$ mass distribution of the process $%
B^{+}\rightarrow D^{+}D^{-}K^{+}$ \cite{LHCb:2020A,LHCb:2020}, enhanced
activity of researches to investigate fully open-flavor exotic mesons. In
fact, by taking into account dominant decays of the resonance-like peaks $%
X_{0(1)}(2900)\rightarrow D^{-}K^{+}$ and assuming that they are four-quark
systems, one sees that $X_{0(1)}(2900)$ are built of quarks $c$, $s$, $u$,
and $d$. The LHCb collaboration measured masses and widths of the structures
$X_{0(1)}(2900)$, and fixed their spin-parities. It was found that $%
X_{0}(2900)$ and $X_{1}(2900)$ bear the quantum numbers $J^{\mathrm{P}%
}=0^{+} $ and $J^{\mathrm{P}}=1^{-}$, respectively. Let us note that,
alternatively, structures $X_{0(1)}(2900)$ may appear as triangle
singularities in some rescattering diagrams: such interpretation was not
excluded by LHCb as well.

In the four-quark picture, widely accepted to explain the LHCb data, $%
X_{0(1)}(2900)$ may be considered in the framework of both the molecule and
diquark-antidiquark (tetraquark) models. Thus, in publications \cite%
{Karliner:2020vsi,Wang:2020xyc} the resonance $X_{0}(2900)$ was analyzed as
the scalar diquark-antidiquark state $[ud][\overline{c}\overline{s}]$,
whereas in Refs.\ \cite{Chen:2020aos} and \cite{Agaev:2020nrc} it was
treated as a molecule $D^{\ast -}K^{\ast +}$ or $\overline{D}^{\ast
0}K^{\ast 0}$. The situation is almost the same for the vector resonance $%
X_{1}(2900)$: it was studied in the context of the tetraquark and molecule
models, for instance, in Refs.\ \cite{Chen:2020aos,Agaev:2021knl,He:2020btl}%
. There are numerous articles devoted to investigations of $X_{0(1)}(2900)$
using different methods and schemes of high energy physics: relatively
complete list of such papers can be found in Refs.\ \cite%
{Agaev:2020nrc,Agaev:2021knl,Agaev:2022eeh}.

An interesting conjecture about nature of $X_{0}(2900)$ was made in Ref.\
\cite{He:2020jna}, where it was interpreted as a radial excitation $%
X_{0}^{\prime }$ of the scalar tetraquark $X_{0}=[ud][\overline{c}\overline{s%
}]$. We addressed this problem in our work \cite{Agaev:2022eeh}, and
calculated masses and widths of the ground-state $1S$ and radially excited $%
2S$ tetraquarks $X_{0}^{(\prime )}$ using the QCD sum rule method. We
modeled $X_{0}^{(\prime )}$ as particles composed of the axial-vector
diquark $[ud]$ and axial-vector antidiquark $[\overline{c}\overline{s}]$. We
also constructed the tetraquarks $X_{\mathrm{S}}^{(\prime )}$ by utilizing a
scalar diquark and antidiquark, and found their parameters. It was
demonstrated that, the ground-state particles $X_{0}$ and $X_{\mathrm{S}}$
are lighter than the resonance $X_{0}(2900)$, whereas radially excited
tetraquarks $X_{0}^{\prime }$ and $X_{\mathrm{S}}^{\prime }$ with the masses
around $\approx 3320\mathrm{\ MeV}$ are heavier it. It other words, none of
these four-quark states can be identified with the resonance $X_{0}(2900)$.
Therefore, it is reasonable to treat $X_{0}$ and $X_{\mathrm{S}}$ as new
hypothetic exotic mesons to be searched for in experiments.

Fully open-flavor tetraquarks, to be fair, were already objects of
theoretical studies, which intensified after information on the resonance $%
X(5568)$ presumably composed of $b$, $s$, $u$, and $d$ quarks \cite%
{D0:2016mwd}. Though existence of $X(5568)$ was not confirmed by other
experimental groups, its charmed partners $b\rightarrow c$ are still under
detailed analysis. In fact, the spectroscopic parameters and full width of
scalar tetraquark $X_{c}=[su][\overline{c}\overline{d}]$ were calculated in
Ref.\ \cite{Agaev:2016lkl}. Masses of exotic mesons with the same content,
but quantum numbers $J^{\mathrm{P}}=0^{+}$ and $J^{\mathrm{P}}=1^{+}$ were
estimated in Ref.\ \cite{Chen:2016mqt}.

In various combinations $c$, $s$, $u$, and $d$ quarks form different classes
of four-quark mesons, features of which deserve investigations. Interesting
class of fully open-flavor particles is collection of states $Z^{++}=[cu][%
\overline{s}\overline{d}]$, which carries two units of electric charge. The
scalar, pseudoscalar, axial-vector and vector members of this group were
studied in our articles Refs.\ \cite{Agaev:2017oay} and \cite{Agaev:2021jsz}%
, respectively. It was pointed out that scalar and vector tetraquarks\ $Z_{%
\mathrm{S}}^{++}$ and $Z_{\mathrm{V}}^{++}$ may be observed in the $%
D^{+}K^{+}$ mass distribution of the decay $B^{+}\rightarrow D^{-}D^{+}K^{+}$
\cite{Agaev:2021jsz}.

Tetraquarks with a content $[ud][\overline{c}\overline{s}]$ establish new
class of open-flavor particles. Observation of the resonances $%
X_{0(1)}(2900) $ by the LHCb collaboration, available experimental data, and
possible interpretation of $X_{1}(2900)$ as a vector state $X_{\mathrm{V}%
}=[ud][\overline{c}\overline{s}]$ make these particles objects of special
interest. In the present article, we continue our studies started in Refs.\
\cite{Agaev:2021knl,Agaev:2022eeh} \ by calculating spectroscopic parameters
and full widths of axial-vector and pseudoscalar four-quark states $X_{%
\mathrm{AV}}$ and $X_{\mathrm{PS}}$ with the same $[ud][\overline{c}%
\overline{s}]$ content.

Masses and current couplings of $X_{\mathrm{AV}}$ and $X_{\mathrm{PS}}$ are
evaluated in the context of the two-point sum rule method \cite%
{Shifman:1978bx,Shifman:1978by}. In calculations, we take into account
various vacuum condensates up to dimension $10$. The full width of the
axial-vector state $X_{\mathrm{AV}}$ is found by including into analysis its
$S$- wave decay modes $X_{\mathrm{AV}}\rightarrow D^{\ast }(2010)^{-}K^{+},$
$\overline{D}^{\ast }(2007)^{0}K^{0}$, $D^{-}K^{\ast }(892)^{+}$, and $%
\overline{D}^{0}K^{\ast }(892)^{0}$. To estimate width of the pseudoscalar
tetraquark $X_{\mathrm{PS}}$, we consider kinematically allowed $S$-wave
channel $X_{\mathrm{PS}}\rightarrow \overline{D}_{0}^{\ast }(2300)^{0}K^{0}$%
, and $P$-wave decay modes $X_{\mathrm{PS}}\rightarrow D^{-}K^{\ast
}(892)^{+}$ and $X_{\mathrm{PS}}\rightarrow D^{0}K^{\ast }(892)^{0}$.

Partial widths of these processes are governed by strong couplings at
relevant vertices, for example, at $X_{\mathrm{AV}}D^{\ast }(2010)^{-}K^{+}$
for the first process. To calculate required couplings, we use the QCD
light-cone sum rule (LCSR) method \cite{Balitsky:1989ry} and soft-meson
approximation \cite{Belyaev:1994zk,Ioffe:1983ju}. The latter is necessary to
treat tetraquark-meson-meson vertices, which due to unequal number of quark
fields in tetraquark and meson interpolating currents differ from standard
three-meson vertices \cite{Agaev:2016dev}.

This paper is organized in the following manner: In Section \ref{sec:Masses}%
, we calculate the masses and current couplings of the tetraquarks $X_{%
\mathrm{AV}}$ and $X_{\mathrm{PS}}$. In Section \ref{sec:Decays1}, we
determine strong couplings $g_{i},\ i=1,2,3,4$ corresponding to vertices $X_{%
\mathrm{AV}}D^{\ast }(2010)^{-}K^{+}$, $X_{\mathrm{AV}}\overline{D}^{\ast
}(2007)^{0}K^{0}$ $X_{\mathrm{AV}}D^{-}K^{\ast }(892)^{+}$, and $X_{\mathrm{%
AV}}\overline{D}^{0}K^{\ast }(892)^{0}$. In this section, we compute partial
widths of corresponding processes, and estimate full width of $X_{\mathrm{AV}%
}$. In Section \ref{sec:Decays2}, we consider the decays $X_{\mathrm{PS}%
}\rightarrow \overline{D}_{0}^{\ast }(2300)^{0}K^{0}$, $D^{-}K^{\ast
}(892)^{+}$, and $D^{0}K^{\ast }(892)^{0}$, and find strong couplings $%
G_{j},\ j=1,2,3$ at relevant vertices. Using $G_{j}$, we calculate partial
width of these decays and evaluate full width of $X_{\mathrm{PS}}$ by
saturating it with these channels. Section \ref{sec:Disc} contains our
conclusions.

%%%%%%%%%%%%%%%%%%%%%%%%%%%%%%%%%%%%%%%%%%%%%%%%%%%%%%%%%%%%%%%%%%%%%%%%%%%%

\section{Masses and current couplings of the tetraquarks $X_{\mathrm{AV}}$
and $X_{\mathrm{PS}}$}

\label{sec:Masses}
%%%%%%%%%%%%%%%%%%%%%%%%%%%%%%%%%%%%%%%%%%%%%%%%%%%%%%%%%%%
In this section, we compute spectroscopic parameters of the states $X_{%
\mathrm{AV}}$ and $X_{\mathrm{PS}}$ by means of the two-point sum rule
method. It is an effective nonperturbative approach elaborated to evaluate
parameters of ordinary mesons and baryons. The QCD sum rules express various
physical quantities in terms of universal vacuum condensates which do not
depend on a problem under consideration. At the same time, they contain
auxiliary parameters $s_{0}$ and $M^{2}$ specific for each computation. The
first of them is the continuum subtraction parameter $s_{0}$ that separates
contribution of a ground-state particle in the phenomenological side of a
sum rule from effects of higher resonances and continuum states. The Borel
parameter $M^{2}$ is required to suppress these unwanted continuum effects.
By introducing $M^{2}$ and $s_{0}$ into analysis and employing an assumption
about quark-hadron duality one connects phenomenological and QCD sides of
sum rules and gets a sum equality. The latter can be used to express
physical observables in terms of different vacuum condensates. The
parameters $M^{2}$ and $s_{0}$ generate theoretical uncertainties in
results, which nevertheless can be estimated and kept under control.

In what follows, we calculate the mass $m$ and current coupling $f$ of the
axial-vector meson $X_{\mathrm{AV}}$ (we employ also $m_{\mathrm{AV}}$ and $%
f_{\mathrm{AV}}$), and provide only final results for $X_{\mathrm{PS}}$. The
starting point in computation of the spectroscopic parameters of the
tetraquark $X_{\mathrm{AV}}$ is the correlation function
\begin{equation}
\Pi _{\mu \nu }(p)=i\int d^{4}xe^{ipx}\langle 0|\mathcal{T}\{J_{\mu
}(x)J_{\nu }^{\dag }(0)\}|0\rangle .  \label{eq:CF1}
\end{equation}%
where, $\mathcal{T}$ is the time-ordered product of two currents, and $%
J_{\mu }(x)$ is the interpolating current for the axial-vector state $X_{%
\mathrm{AV}}$. We model the tetraquark $X_{\mathrm{AV}}$ as a compound
formed by the scalar diquark $u^{T}C\gamma _{5}d$ and axial-vector
antidiquark $\overline{c}\gamma _{\mu }C\overline{s}^{T}$, which are
antitriplet and triplet states of the color group $SU_{c}(3)$, respectively.
Therefore, corresponding interpolating current is given by the formula
\begin{equation}
J_{\mu }(x)=\epsilon \widetilde{\epsilon }u_{b}^{T}(x)C\gamma _{5}d_{c}(x)%
\overline{c}_{d}(x)\gamma _{\mu }C\overline{s}_{e}^{T}(x),  \label{eq:CR1}
\end{equation}%
and belongs to $[\overline{\mathbf{3}}_{c}]_{ud}\otimes \lbrack \mathbf{3}%
_{c}]_{\overline{c}\overline{s}}$ representation of the color group. In
expression above, $\epsilon \widetilde{\epsilon }=\epsilon _{abc}\epsilon
_{ade}$, where $a$, $b$, $c$, $d$ and $e$ are color indices. In Eq.\ (\ref%
{eq:CR1}) $c(x)$, $s(x)$, $u(x)$ and $d(x)$ denote quark fields, and $C$ is
the charge conjugation matrix.

The phenomenological side of the sum rule $\Pi _{\mu \nu }^{\mathrm{Phys}%
}(p) $
\begin{equation}
\Pi _{\mu \nu }^{\mathrm{Phys}}(p)=\frac{\langle 0|J_{\mu }|X_{\mathrm{AV}%
}(p,\varepsilon )\rangle \langle X_{\mathrm{AV}}(p,\varepsilon )|J_{\nu
}^{\dagger }|0\rangle }{m^{2}-p^{2}}+\cdots,  \label{eq:Phys1}
\end{equation}%
is derived from Eq.\ (\ref{eq:CF1}) by inserting a complete set of
intermediate states with quark contents and spin-parity of the tetraquark $%
X_{\mathrm{AV}}$, and carrying out integration over $x$. The momentum and
polarization vector of $X_{\mathrm{AV}}$ are denoted by $p$ and $\varepsilon
$, respectively. It should be noted that in $\Pi _{\mu \nu }^{\mathrm{Phys}%
}(p)$ the ground-state term is written down explicitly, whereas
contributions of higher resonances and continuum states are shown by
ellipses.

In Eq.\ (\ref{eq:Phys1}), we have assumed that the phenomenological side of
the sum rule $\Pi _{\mu \nu }^{\mathrm{Phys}}(p)$ can be approximated by a
single pole term. But in the case of multiquark systems this approximation
has to be used with some caution, because the physical side receives
contribution also from two-hadron reducible terms. Indeed, a relevant
interpolating current couples not only to a multiquark hadron, but also to a
two-hadron continuum. This problem was raised in Refs.\ \cite%
{Kondo:2004cr,Lee:2004xk} when considering pentaquarks, and revisited
recently in the case of tetraquarks \cite{Lucha:2019pmp}, where it is argued that the contributions at the orders ${\cal O} (1) $ and ${\cal O} (\alpha_s) $  in the operator product expansion  (OPE)  are canceled out exactly  by the meson-meson scattering states at the hadronic side and the tetraquark molecular states start to receive contributions at the order ${\cal O }(\alpha^2_s)$. Then the reducible
contributions should be subtracted from the sum rule, which can be done by
means of two methods. One of them is direct subtraction of two-hadron terms
from $\Pi _{\mu \nu }^{\mathrm{Phys}}(p)$ by calculating current-two-hadron
coupling constant using an independent QCD sum rule. This strategy was
realized, for example, in Ref.\ \cite{Sarac:2005fn} to investigate
anti-charmed pentaquark state. Existence of a two-hadron continuum below a
multiquark system means that such particle is unstable and decays to these
conventional hadrons. In other words, a two-hadron continuum generates the
finite width $\Gamma (p^{2})$ of a multiquark system. Relevant effects can
be taken into account by modifying the quark propagator in Eq.\ (\ref%
{eq:Phys1})
\begin{equation}
\frac{1}{m^{2}-p^{2}}\rightarrow \frac{1}{m^{2}-p^{2}-i\sqrt{p^{2}}\Gamma (p)%
}.  \label{eq:Mod}
\end{equation}%
This second method was used to study the tetraquarks \cite{Wang:2015nwa}.
Rather detailed investigations demonstrated that effects of the modification
Eq.\ (\ref{eq:Mod}) can be taken into account by absorbing two-meson
contributions into a current-tetraquark coupling constant and keeping stable
tetraquark's mass \cite{Agaev:2018vag,Sundu:2018nxt}. Uncertainties
generated by changing of a coupling are numerically smaller than theoretical
errors of sum rule analysis itself. In fact, two-meson effects lead to
additional $\approx 7\%$ uncertainty in the current coupling $f_{T}$\ for
doubly charmed pseudoscalar tetraquark $cc\overline{s}\overline{s}$ with the
mass $m_{T}=4390~\mathrm{MeV}$ and full width $\Gamma _{T}\approx 300~%
\mathrm{MeV}$ \cite{Agaev:2018vag}. In the case of the resonance $%
Z_{c}^{-}(4100)$ these uncertainties do not exceed $\approx 5\%$ of the
coupling $f_{Z_{c}}$ \cite{Sundu:2018nxt}. Therefore, one can neglect
two-meson reducible terms and use in $\Pi _{\mu \nu }^{\mathrm{Phys}}(p)$
single-pole zero-width approximation, as it has been done in Eq.\ (\ref%
{eq:Phys1}).

To simplify the correlation function $\Pi _{\mu \nu }^{\mathrm{Phys}}(p)$
and express it in terms of the tetraquark's mass and current coupling, we
use the matrix element
\begin{equation}
\langle 0|J_{\mu }|X_{\mathrm{AV}}(p,\varepsilon )\rangle =fm\varepsilon
_{\mu },  \label{eq:ME1}
\end{equation}%
and recast $\Pi _{\mu \nu }^{\mathrm{Phys}}(p)$ into the following form
\begin{equation}
\Pi _{\mu \nu }^{\mathrm{Phys}}(p)=\frac{f^{2}m^{2}}{m^{2}-p^{2}}\left(
-g_{\mu \nu }+\frac{p_{\mu }p_{\nu }}{m^{2}}\right) +\cdots .
\label{eq:Phen2}
\end{equation}%
The function $\Pi _{\mu \nu }^{\mathrm{Phys}}(p)$ has two Lorentz structures
determined by $g_{\mu \nu }$ and $p_{\mu }p_{\nu }$. One of them can be
chosen to continue sum rule analysis. We work with the structure
proportional to $g_{\mu \nu }$ and corresponding invariant amplitude $\Pi ^{%
\mathrm{Phys}}(p^{2})$. Advantage of this structure is that it is formed due
to contributions of only spin-1 particles, and is free of any contaminations.

The QCD side of the sum rules $\Pi _{\mu \nu }^{\mathrm{OPE}}(p)$ should be
computed in the operator product expansion with some accuracy. To this end,
we substitute into $\Pi _{\mu \nu }(p)$ explicit expression of the current $%
J_{\mu }(x)$, contract relevant quark fields, and replace contractions by
appropriate propagators. These operations lead to the expression
\begin{eqnarray}
&&\Pi _{\mu \nu }^{\mathrm{OPE}}(p)=i\int d^{4}xe^{ipx}\epsilon \widetilde{%
\epsilon }\epsilon ^{\prime }\widetilde{\epsilon }^{\prime }\mathrm{Tr}\left[
\gamma _{5}\widetilde{S}_{u}^{bb^{\prime }}(x)\right.  \notag \\
&&\times \left. \gamma _{5}S_{d}^{cc^{\prime }}(x)\right] \mathrm{Tr}\left[
\gamma _{\mu }\widetilde{S}_{s}^{e^{\prime }e}(-x)\gamma _{\nu
}S_{c}^{d^{\prime }d}(-x)\right] ,  \label{eq:QCD1}
\end{eqnarray}%
where%
\begin{equation}
\widetilde{S}_{q}(x)=CS_{q}^{T}(x)C,  \label{eq:Prop}
\end{equation}%
and $\epsilon ^{\prime }\widetilde{\epsilon }^{\prime }=\epsilon _{a^{\prime
}b^{\prime }c^{\prime }}\epsilon _{a^{\prime }d^{\prime }e^{\prime }}$.
Here, $S_{c}(x)$ and $S_{u(s,d)}(x)$ are the heavy $c$- and light $u(s,d)$%
-quark propagators, respectively: Their explicit expressions are presented
in Appendix (see, also Ref.\ \cite{Agaev:2020zad}).

The function $\Pi _{\mu \nu }^{\mathrm{OPE}}(p)$ is a sum of components
proportional to $g_{\mu \nu }$ and $p_{\mu }p_{\nu }$. We choose the
invariant amplitude $\Pi ^{\mathrm{OPE}}(p^{2})$ corresponding to structure $%
\sim g_{\mu \nu }$, and use it to derive sum rules for $m$ and $f$ , which
read
\begin{equation}
m^{2}=\frac{\Pi ^{\prime }(M^{2},s_{0})}{\Pi (M^{2},s_{0})},  \label{eq:Mass}
\end{equation}%
and
\begin{equation}
f^{2}=\frac{e^{m^{2}/M^{2}}}{m^{2}}\Pi (M^{2},s_{0}).  \label{eq:Coupl}
\end{equation}%
In expressions above, $\Pi (M^{2},s_{0})$ is the Borel transformed and
subtracted invariant amplitude $\Pi ^{\mathrm{OPE}}(p^{2})$, and $\Pi
^{\prime }(M^{2},s_{0})=d\Pi (M^{2},s_{0})/d(-1/M^{2})$.

Computing the function $\Pi (M^{2},s_{0})$ and fixing of regions for
parameters $M^{2}$ and $s_{0}$ are next problems in our study of $m$ and $f$%
. Calculations prove that $\Pi (M^{2},s_{0})$ has the form%
\begin{equation}
\Pi (M^{2},s_{0})=\int_{\mathcal{M}^{2}}^{s_{0}}ds\rho ^{\mathrm{OPE}%
}(s)e^{-s/M^{2}}+\Pi (M^{2}),  \label{eq:InvAmp}
\end{equation}%
where $\mathcal{M}=m_{c}+m_{s}$. In the present paper, we neglect the masses
of $u$ and $d$ quarks, as well as set $m_{s}^{2}=0$ saving, at the same
time, terms $\sim m_{s}$. The spectral density $\rho ^{\mathrm{OPE}}(s)$ is
found as an imaginary part of the function $\Pi ^{\mathrm{OPE}}(p^{2})$.
Borel transformation some of terms are computed directly from expression of $%
\Pi _{\mu \nu }^{\mathrm{OPE}}(p)$: They form the second component $\Pi
(M^{2})$ in Eq.\ (\ref{eq:InvAmp}). Our analysis includes contributions of
different quark, gluon and mixed vacuum condensates up to dimension $10$.
Full analytical expressions of $\rho ^{\mathrm{OPE}}(s)$ and $\Pi (M^{2})$
are written down in Appendix.

The vacuum condensates, that enter to sum rules Eqs.\ (\ref{eq:Mass}) and (%
\ref{eq:Coupl}), are universal quantities obtained from analysis of various
hadronic processes \cite%
{Shifman:1978bx,Shifman:1978by,Ioffe:1981kw,Ioffe:2005ym}. Below, we list
their values used in our numerical computations
\begin{eqnarray}
&&\langle \overline{q}q\rangle =-(0.24\pm 0.01)^{3}~\mathrm{GeV}^{3},\
\langle \overline{s}s\rangle =(0.8\pm 0.01)\langle \overline{q}q\rangle ,
\notag \\
&&\langle \overline{q}g_{s}\sigma Gq\rangle =m_{0}^{2}\langle \overline{q}%
q\rangle ,\ \langle \overline{s}g_{s}\sigma Gs\rangle =m_{0}^{2}\langle
\overline{s}s\rangle ,  \notag \\
&&m_{0}^{2}=(0.8\pm 0.2)~\mathrm{GeV}^{2},  \notag \\
&&\langle \frac{\alpha _{s}G^{2}}{\pi }\rangle =(0.012\pm 0.004)~\mathrm{GeV}%
^{4},  \notag \\
&&\langle g_{s}^{3}G^{3}\rangle =(0.57\pm 0.29)~\mathrm{GeV}^{6},\
m_{c}=1.27\pm 0.02~\mathrm{GeV},  \notag \\
&&m_{s}=93_{-5}^{+11}~\mathrm{MeV}.  \label{eq:Parameters}
\end{eqnarray}%
As is seen, the vacuum condensate of strange quarks differs from $\langle 0|%
\overline{q}q|0\rangle $ \cite{Ioffe:1981kw}. The mixed condensates $\langle
\overline{q}g_{s}\sigma Gq\rangle $ and $\langle \overline{s}g_{s}\sigma
Gs\rangle $ are expressed using the corresponding quark condensates and
parameter $m_{0}^{2}$, numerical value of which was extracted from analysis
of baryonic resonances \cite{Ioffe:2005ym}. For the gluon condensate $%
\langle g^{3}G^{3}\rangle $, we employ the estimate given in Ref.\ \cite%
{Narison:2015nxh}. This list also contains the masses of $c$ and $s$ quarks
from Ref.\ \cite{PDG:2020} in the $\overline{\mathrm{MS}}$-scheme.

\begin{figure}[h]
\includegraphics[width=8cm]{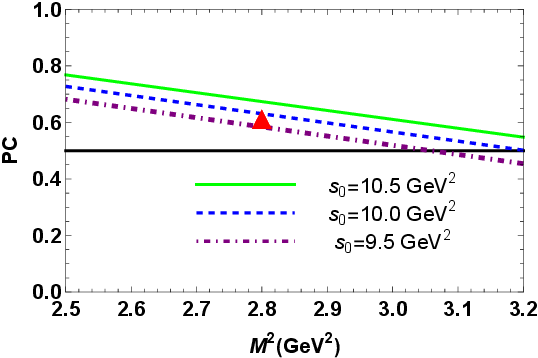}
\caption{The pole contribution $\mathrm{PC}$ as a function of the Borel
parameter $M^{2}$ at different $s_{0}$. The horizontal black line limits a
region $\mathrm{PC}=0.5$. The red triangle marks the point, where the mass $%
m $ of $X_{\mathrm{AV}}$ has effectively been computed. }
\label{fig:PC}
\end{figure}

\begin{figure}[h]
\includegraphics[width=8cm]{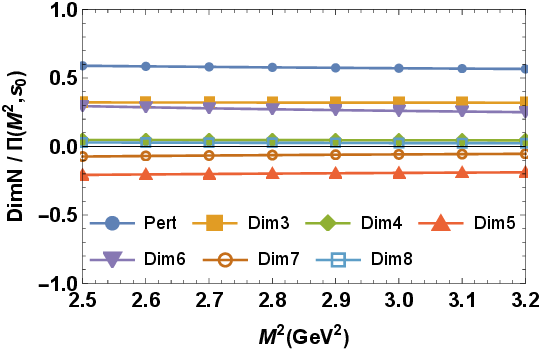}
\caption{Different contributions to $\Pi (M^{2},s_{0})$ as functions of $M^2$%
. Dimension-$9$ and $10$ terms are very small and not shown in the plot. All
curves have been calculated at $s_0=10~\mathrm{GeV}^2$.}
\label{fig:Conv}
\end{figure}

\begin{widetext}

\begin{figure}[h!]
\begin{center}
\includegraphics[totalheight=6cm,width=8cm]{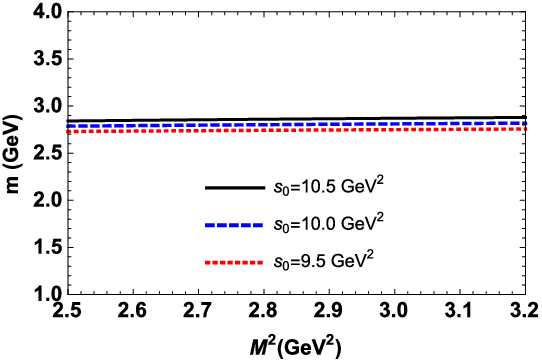}
\includegraphics[totalheight=6cm,width=8cm]{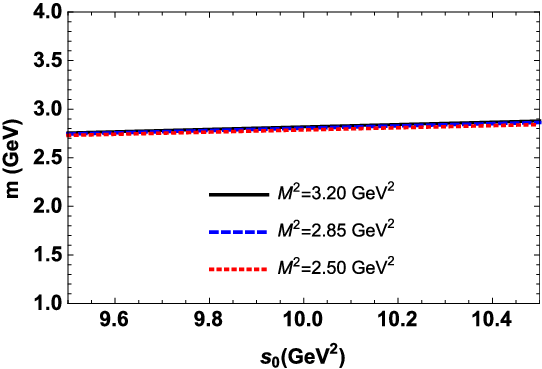}
\end{center}
\caption{Mass $m$ of the tetraquark $X_{\mathrm{AV}}$ as a function of the Borel parameter $M^{2}$ (left), and the continuum threshold parameter $s_0$ (right).}
\label{fig:Mass}
\end{figure}

\end{widetext}

Another problem is a choice of working windows for parameters $M^{2}$ and $%
s_{0}$. They are fixed in such a way that to meet constraints imposed on $%
\Pi (M^{2},s_{0})$ by a pole contribution ($\mathrm{PC}$) and convergence of
the operator product expansion. These constraints can be quantified by means
of the following expressions%
\begin{equation}
\mathrm{PC}=\frac{\Pi (M^{2},s_{0})}{\Pi (M^{2},\infty )},  \label{eq:PC}
\end{equation}%
and
\begin{equation}
R(M^{2})=\frac{\Pi ^{\mathrm{DimN}}(M^{2},s_{0})}{\Pi (M^{2},s_{0})}.
\label{eq:Convergence}
\end{equation}%
where $\Pi ^{\mathrm{DimN}}(M^{2},s_{0})$ is a sum of $\mathrm{DimN\equiv
Dim(8+9+10)}$ terms. In what follows, we require fulfilment of the
restrictions
\begin{equation}
\mathrm{PC}\geq 0.5\text{, }R(M^{2})\leq 0.05.  \label{eq:Constr}
\end{equation}

The $\mathrm{PC}$ and $R(M^{2})$ are employed to fix the higher and lower
limits of the Borel parameter $M^{2}$, respectively. These two values
determine boundaries of the region where $M^{2}$ can be varied. Calculations
show that intervals
\begin{equation}
M^{2}\in \lbrack 2.5,3.2]\ \mathrm{GeV}^{2},\ s_{0}\in \lbrack 9.5,10.5]\
\mathrm{GeV}^{2},  \label{eq:Wind1}
\end{equation}%
are appropriate regions for the parameters $M^{2}$ and $s_{0}$, and comply
with limits on $\mathrm{PC}$ and convergence of $\mathrm{OPE}$. Thus, at $M_{%
\mathrm{max}}^{2}=3.2~\mathrm{GeV}^{2}$ on average in $s_{0}$ the pole
contribution is $0.51$, whereas at $M_{\mathrm{min}}^{2}=2.5~\mathrm{GeV}%
^{2} $ it becomes equal to $0.73$. To visualize dynamics of the pole
contribution when varying the Borel parameter, we plot $\mathrm{PC}$ as a
function of $M^{2}$ at different $s_{0}$ in Fig.\ \ref{fig:PC}. One can see,
that except for a small region $M^{2}\geq 3.1~\mathrm{GeV}^{2}$ at $%
s_{0}=9.5~\mathrm{GeV}^{2}$ the pole contribution exceeds $0.5$. On average
in $s_{0}$, the condition $\mathrm{PC}\geq 0.5$ is fulfilled in the whole
working region Eq.\ (\ref{eq:Wind1}).

To be convinced in convergence of $\mathrm{OPE}$, we calculate $\ R(M_{%
\mathrm{min}}^{2})$ at the minimum point $M_{\mathrm{min}}^{2}=2.5~\mathrm{%
GeV}^{2}$, and get $R(2.5~\mathrm{GeV}^{2})\approx 0.027$ in accordance with
constraint from Eq.\ (\ref{eq:Constr}). Results of more detailed analysis
are depicted in Fig.\ \ref{fig:Conv}. In this figure, we show the
perturbative and nonperturbative components of the correlation function $\Pi
(M^{2},s_{0})$: A prevalence of the perturbative contribution to $\Pi
(M^{2},s_{0})$ over nonperturbative one is evident. Without regard for some
higher dimensional terms, the nonperturbative contributions reduce by
increasing the dimensions of the corresponding operators.

The region for $s_{0}$ has to meet constraints coming from dominance of $%
\mathrm{PC}$ and convergence of $\mathrm{OPE}$. Self-consistency of
performed analysis can be checked by comparing the parameter $\sqrt{s_{0}}$
and the $X_{\mathrm{AV}}$ tetraquark's mass extracted from the sum rule:
Evidently, an inequality $m<\sqrt{s_{0}}$ should be satisfied.
Additionally, $\sqrt{s_{0}}$ bears information on a mass $m^{\ast }$ of the
first radial excitation of the tetraquark $X_{\mathrm{AV}}$, therefore the
restriction $m^{\ast }\geq $ $\sqrt{s_{0}}$ provides low limit for $m^{\ast }
$.

We extract the mass $m$ and coupling $f$ by computing them at different $%
M^{2}$ and $s_{0}$, and finding their mean values averaged over the regions
Eq.\ (\ref{eq:Wind1}). Our predictions for $m$ and $f$ read
\begin{eqnarray}
m &=&(2800~\pm 75)~\mathrm{MeV},  \notag \\
f &=&(2.42\pm 0.30)\times 10^{-3}~\mathrm{GeV}^{4}.  \label{eq:Result1}
\end{eqnarray}%
The results in Eq.\ (\ref{eq:Result1}) effectively correspond to sum rules'
predictions at approximately middle point of the regions in Eq.\ (\ref%
{eq:Wind1}), i.e., to predictions at the point $M^{2}=2.8~\mathrm{GeV}^{2}$
and $s_{0}=10~\mathrm{GeV}^{2}$, where the pole contribution is $\mathrm{PC}%
\approx 0.62$. This fact guarantees a dominance of the pole contribution in
extracted parameters $m$ and $f$.

\begin{widetext}

\begin{figure}[h!]
\begin{center}
\includegraphics[totalheight=6cm,width=8cm]{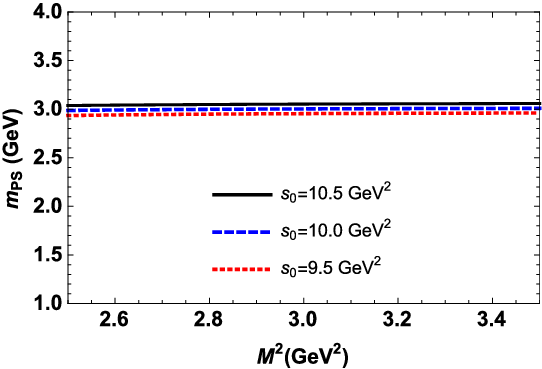}
\includegraphics[totalheight=6cm,width=8cm]{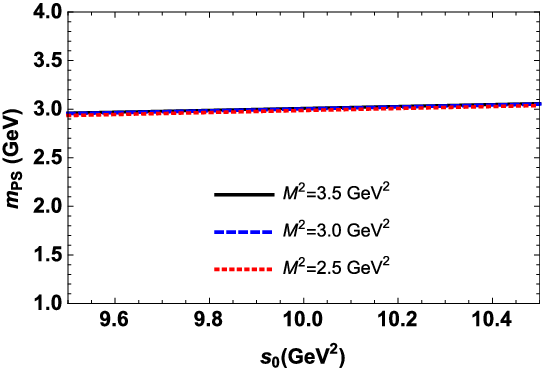}
\end{center}
\caption{Dependence of the mass $m_{\mathrm{PS}}$ on the Borel parameter $M^{2}$ (left), and on the continuum threshold parameter $s_0$ (right).}
\label{fig:Mass2}
\end{figure}

\end{widetext}

In Fig.\ \ref{fig:Mass}, we depict $m$ as functions of $M^{2}$ and $s_{0}$,
in which is seen its dependence on the Borel and continuum subtraction
parameters. Strictly speaking, physical quantities should not depend on $%
M^{2}$, but computations demonstrate that such effects, nevertheless, exist.
Therefore, in a chosen region for $M^{2}$ this dependence should be minimal.
Due to a functional form of the sum rule for the mass Eq.\ (\ref{eq:Mass})
given as the ratio of correlation functions, variation of $m$ in the region
for $M^{2}$ is mild. There is also dependence on the parameter $s_{0}$ which
contains information about the lower limit for the mass of the excited
tetraquark. 
%Indeed, it is known that $s_{0}$ dissects a ground-state
%contribution from ones of higher resonances and continuum states. Hence, $%
%s_{0}$ fixed in accordance with constraints imposed on $\mathrm{PC}$ and $%
%\mathrm{OPE}$, provides lower limit for the mass $m^{\prime }$ of the
%excited state $\sqrt{s_{0}}\leq m^{\prime }$.

The pseudoscalar tetraquark $X_{\mathrm{PS}}$ and its parameters have been
explored by the manner described just above. Here, we model $X_{\mathrm{PS}}$
as a tetraquark built of the pseudoscalar diquark $u^{T}Cd$ and scalar
antidiquark $\overline{c}\gamma _{5}C\overline{s}^{T}$. The relevant
interpolating current $J_{\mathrm{PS}}(x)$ is determined by the expression
\begin{equation}
J_{\mathrm{PS}}(x)=\epsilon \widetilde{\epsilon }u_{b}^{T}(x)Cd_{c}(x)%
\overline{c}_{d}(x)\gamma _{5}C\overline{s}_{e}^{T}(x),  \label{eq:CR2}
\end{equation}%
and belongs to the antitriplet-triplet representation of the color group $%
SU_{c}(3)$.

The physical side of the sum rule in this case has relatively simple form%
\begin{equation}
\widetilde{\Pi }^{\mathrm{Phys}}(p)=\frac{f_{\mathrm{PS}}^{2}m_{\mathrm{PS}%
}^{4}}{(m_{c}+m_{s})^{2}(m_{\mathrm{PS}}^{2}-p^{2})}+\cdots ,
\label{eq:Phys2}
\end{equation}%
where $m_{\mathrm{PS}}$ and $f_{\mathrm{PS}}$ are the mass and current
coupling of the tetraquark $X_{\mathrm{PS}}$, respectively. To derive $%
\widetilde{\Pi }^{\mathrm{Phys}}(p)$, we have used the matrix element of the
pseudoscalar particle $X_{\mathrm{PS}}$
\begin{equation}
\langle 0|J_{\mathrm{PS}}|X_{\mathrm{PS}}(p)\rangle =\frac{f_{\mathrm{PS}}m_{%
\mathrm{PS}}^{2}}{m_{c}+m_{s}}.  \label{eq:Mel1a}
\end{equation}%
The function $\widetilde{\Pi }^{\mathrm{Phys}}(p)$ has trivial Lorentz
structure proportional to $I$, therefore the invariant amplitude $\widetilde{%
\Pi }^{\mathrm{Phys}}(p^{2})$ is equal to r.h.s. of Eq.\ (\ref{eq:Phys2}).

The QCD side of new sum rules is given by the formula
\begin{eqnarray}
&&\widetilde{\Pi }^{\mathrm{OPE}}(p)=i\int d^{4}xe^{ipx}\epsilon \widetilde{%
\epsilon }\epsilon ^{\prime }\widetilde{\epsilon }^{\prime }\mathrm{Tr}\left[
\widetilde{S}_{u}^{bb^{\prime }}(x)S_{d}^{cc^{\prime }}(x)\right]  \notag \\
&&\times \mathrm{Tr}\left[ \gamma _{5}\widetilde{S}_{s}^{e^{\prime
}e}(-x)\gamma _{5}S_{c}^{d^{\prime }d}(-x)\right] .  \label{eq:QCD2}
\end{eqnarray}%
The spectroscopic parameters of $X_{\mathrm{PS}}$ can be obtained from Eqs.\
(\ref{eq:Mass}) and\ (\ref{eq:Coupl}) after replacement $\Pi
(M^{2},s_{0})\rightarrow \widetilde{\Pi }(M^{2},s_{0})$. Performed
calculations yield%
\begin{eqnarray}
m_{\mathrm{PS}} &=&(3000\pm 60)~\mathrm{MeV},  \notag \\
f_{\mathrm{PS}} &=&(8.69\pm 1.54)\times 10^{-4}~\mathrm{GeV}^{4}.
\label{eq:Result2}
\end{eqnarray}%
The Borel and continuum subtraction parameters $M^{2}$ and $s_{0}$ used to
extract $m_{\mathrm{PS}}$ and $f_{\mathrm{PS}}$ are given by Eq.\ (\ref%
{eq:Wind2})%
\begin{equation}
M^{2}\in \lbrack 2.5,3.5]\ \mathrm{GeV}^{2},\ s_{0}\in \lbrack 9.5,10.5]\
\mathrm{GeV}^{2}.  \label{eq:Wind2}
\end{equation}
In these regions the $\mathrm{PC}$ changes inside limits%
\begin{equation}
0.74\geq \mathrm{PC}\geq 0.50.
\end{equation}%
Dependence of $m_{\mathrm{PS}}$ on the parameters $M^{2}$ and $s_{0}$ is
shown in Fig.\ \ref{fig:Mass2}. In the left panel one can see a relatively
stable nature of $m_{\mathrm{PS}}$ under variation of $M^{2}$.

%%%%%%%%%%%%%%%%%%%%%%%%%%%%%%%%%%%%%%%%%%%%%%%%%%%%%%%%%%%%%%%%%%%%%%%%%%%%

\section{Decays of the axial-vector tetraquark $X_{\mathrm{AV}}$}

\label{sec:Decays1}
%%%%%%%%%%%%%%%%%%%%%%%%%%%%%%%%%%%%%%%%%%%%%%%%%%%%%%%%%%%

The mass and spin-parity of the tetraquark $X_{\mathrm{AV}}$ allow us to
classify its decay channels. We restrict ourselves by analysis of $S$-wave
decay channels of $X_{\mathrm{AV}}$ which are $X_{\mathrm{AV}}\rightarrow
D^{\ast }(2010)^{-}K^{+}$, $\overline{D}^{\ast }(2007)^{0}K^{0}$, $%
D^{-}K^{\ast }(892)^{+}$, and $\overline{D}^{0}K^{\ast }(892)^{0}$. The full
width of the axial-vector state $X_{\mathrm{AV}}$ is estimated by including
into analysis namely these channels.

We are going to provide rather detailed information about computation of a
partial width of the decay $X_{\mathrm{AV}}\rightarrow D^{\ast
}(2010)^{-}K^{+}$, and outline important steps in analyses of other
processes. A quantity to be extracted from a sum rule is the strong coupling
$g_{1}$ of particles at the vertex $X_{\mathrm{AV}}D^{\ast }(2010)^{-}K^{+}$. 
This coupling is defined in terms of the on-mass-shell matrix element
\begin{eqnarray}
\langle D^{\ast }\left( p\right) K(q)|X_{\mathrm{AV}}(p^{\prime })\rangle
&=&g_{1}\left[ \left( p\cdot p^{\prime }\right) \left( \varepsilon ^{\ast
}\cdot \varepsilon ^{\prime }\right) \right.  \notag \\
&&\left. -\left( p\cdot \varepsilon ^{\prime }\right) \left( p^{\prime
}\cdot \varepsilon ^{\ast }\right) \right] ,  \label{eq:Mel2}
\end{eqnarray}%
where the mesons $K^{+}$ and $D^{\ast }(2010)^{-}$ are denoted as $K$ and $%
D^{\ast }$, respectively. Here, $p^{\prime }$, $p$ and $q$ are four-momenta
of the tetraquark $X_{\mathrm{AV}}$, and mesons $D^{\ast }$ and $K$, and $%
\varepsilon _{\nu }^{\prime }$ and $\varepsilon _{\mu }^{\ast }$ are the
polarization vectors of the particles $X_{\mathrm{AV}}$ and $D^{\ast }$.

In the framework of the LCSR method the coupling $g_{1}$ can be obtained
from the correlation function
\begin{equation}
\Pi _{\mu \nu }(p,q)=i\int d^{4}xe^{ipx}\langle K(q)|\mathcal{T}\{J_{\mu
}^{D^{\ast }}(x)J_{\nu }^{\dag }(0)\}|0\rangle ,  \label{eq:CorrF3}
\end{equation}%
with $J_{\nu }(x)$ being the current for the tetraquark $X_{\mathrm{AV}}$
from Eq.\ (\ref{eq:CR1}). The interpolating current for the meson $D^{\ast
}(2010)^{-}$ is abbreviated in Eq.\ (\ref{eq:CorrF3}) as $J_{\mu }^{D^{\ast
}}(x)$, and defined by the expression
\begin{equation}
J_{\mu }^{D^{\ast }}(x)=\overline{c}_{j}(x)\gamma _{\mu }d_{j}(x),
\label{eq:Dcur}
\end{equation}%
where $j$ is the color index.

The main contribution to the correlation function $\Pi _{\mu \nu }(p,q)$
comes from a term with poles at $p^{2}$ and $p^{\prime 2}=(p+q)^{2}$. This
term is given by the formula
\begin{eqnarray}
&&\Pi _{\mu \nu }^{\mathrm{Phys}}(p,q)=g_{1}\frac{fmf_{D^{\ast }}m_{D^{\ast
}}}{\left( p^{2}-m_{D^{\ast }}^{2}\right) \left( p^{\prime 2}-m^{2}\right) }
\notag \\
&&\times \left( \frac{m^{2}+m_{D^{\ast }}^{2}-m_{K}^{2}}{2}g_{\mu \nu
}-p_{\mu }p_{\nu }^{\prime }\right) +\cdots ,  \label{eq:CorrF5}
\end{eqnarray}%
where $m_{D^{\ast }}$ and $f_{D^{\ast }}$ are the mass and decay constant of
the meson $D^{\ast }(2010)^{-}$. To derive Eq.\ (\ref{eq:CorrF5}), we have
used Eq.\ (\ref{eq:Mel2}) and the following matrix elements
\begin{equation}
\langle 0|J_{\mu }^{D^{\ast }}|D^{\ast }(p)\rangle =f_{D^{\ast }}m_{D^{\ast
}}\varepsilon _{\mu },\ \langle X_{\mathrm{AV}}(p^{\prime })|J_{\nu
}^{\dagger }|0\rangle =fm\varepsilon _{\nu }^{\prime \ast }.  \label{eq:Mel1}
\end{equation}%
The term written down explicitly in Eq.\ (\ref{eq:CorrF5}) corresponds to
contribution of ground-state particles in $X_{\mathrm{AV}}$ and $D^{\ast }$
channels: effects of higher resonances and continuum states in these
channels are shown by dots. The function $\Pi _{\mu \nu }^{\mathrm{Phys}%
}(p,q)$ constitutes the phenomenological side of a sum rule for the coupling
$g_{1}$. It contains two terms determined by structures $g_{\mu \nu }$ and $%
p_{\mu }p_{\nu }^{\prime }$. In our studies, we use the term $\sim g_{\mu
\nu }$ and corresponding invariant amplitude $\Pi ^{\mathrm{Phys}%
}(p^{2},p^{\prime 2})$ which is a function of two variables $p^{2}$ and $%
p^{\prime 2}$.

The correlation function $\Pi _{\mu \nu }(p,q)$ calculated in terms of
quark-gluon degrees of freedom forms the QCD side of the sum rules and is
equal to
\begin{eqnarray}
&&\Pi _{\mu \nu }^{\mathrm{OPE}}(p,q)=i\int d^{4}xe^{ipx}\epsilon \widetilde{%
\epsilon }\left[ \gamma _{5}\widetilde{S}_{d}^{jc}(x){}\gamma ^{\mu }\right.
\notag \\
&&\left. \times \widetilde{S}_{c}^{dj}(-x){}\gamma _{\nu }\right] _{\alpha
\beta }\langle K(q)|\overline{u}_{\alpha }^{b}(0)s_{\beta }^{e}(0)|0\rangle ,
\label{eq:CorrF6}
\end{eqnarray}%
where $\alpha $ and $\beta $ are the spinor indices.

As is seen, besides $c$ and $d$ quark propagators the function $\Pi_{\mu\nu} ^{%
\mathrm{OPE}}(p,q)$ contains also local matrix elements of the $K^{+}$
meson, which carry spinor and color indices. We can rewrite $\langle K|%
\overline{u}_{\alpha }^{b}s_{\beta }^{e}|0\rangle $ in convenient forms by
expanding $\overline{u}s$ over the full set of Dirac matrices $\Gamma ^{J}$
\begin{equation}
\Gamma ^{J}=\mathbf{1},\ \gamma _{5},\ \gamma _{\mu },\ i\gamma _{5}\gamma
_{\mu },\ \sigma _{\mu \nu }/\sqrt{2},  \label{eq:Dirac}
\end{equation}%
and projecting them onto the colorless states
\begin{equation}
\overline{u}_{\alpha }^{b}(0)s_{\beta }^{a}(0)\rightarrow \frac{1}{12}\delta
^{ba}\Gamma _{\beta \alpha }^{J}\left[ \overline{u}(0)\Gamma ^{J}s(0)\right]
.  \label{eq:MatEx}
\end{equation}%
Operators $\overline{u}\Gamma ^{J}s$ sandwiched between the $K$ meson and
vacuum generate local matrix elements of the $K$ meson, which are known and
can be implemented into $\Pi _{\mu \nu }^{\mathrm{OPE}}(p,q)$.

As usual, $\Pi _{\mu \nu }^{\mathrm{OPE}}(p,q)$-type correlators depend on
non-local matrix elements of a final meson (for example, $K$ meson), which
are convertible to its distribution amplitudes (DAs). This is correct while
one treats strong vertices of three conventional mesons in the context of
the LCSR method. In the case of tetraquark-meson-meson vertices relevant
correlation functions instead of DAs of a final meson contain its local
matrix elements. These matrix elements are determined at the space-time
point $x=0$ and are overall normalization factors. Within the LCSR method
similar behavior of correlation functions was seen in a limit $q\rightarrow
0 $ of three-meson vertices, which is known as a soft-meson approximation
\cite{Belyaev:1994zk}. This approximation requires adoption of additional
technical tools to deal with new problems appeared in a phenomenological
side of corresponding sum rules \cite{Belyaev:1994zk,Ioffe:1983ju}. It turns
out that the soft limit and related technical methods can be adapted to
investigate also tetraquark-meson-meson vertices \cite{Agaev:2016dev}. It is
worth to emphasize that the soft limit should be implemented in a hard part
of the correlation function $\Pi _{\mu \nu }(p)$, but in matrix elements one
takes into account the terms with $q^{2}=m_{K}^{2}$.

The term proportional to $g_{\mu \nu }$ in $\Pi _{\mu \nu }^{\mathrm{Phys}%
}(p,q)$ in the limit $q\rightarrow 0$ with some accuracy can be transformed
into the expression
\begin{equation}
\Pi _{\mu \nu }^{\mathrm{Phys}}(p)=g_{1}\frac{fmf_{D^{\ast }}m_{D^{\ast }}}{%
\left( p^{2}-\overline{m}^{2}\right) ^{2}}\left( \overline{m}^{2}-\frac{%
m_{K}^{2}}{2}\right) g_{\mu \nu }+\cdots ,  \label{eq:CorrF5a}
\end{equation}%
where $\overline{m}^{2}=(m^{2}+m_{D^{\ast }}^{2})/2$. The invariant
amplitude $\Pi ^{\mathrm{Phys}}(p^{2})$ depends on the variable $p^{2}$, and
has a double pole at $p^{2}=\overline{m}^{2}$. The Borel transformation of $%
\Pi ^{\mathrm{Phys}}(p^{2})$ is given by the formula
\begin{equation}
\mathcal{B}\Pi ^{\mathrm{Phys}}(p^{2})=g_{1}fmf_{D^{\ast }}m_{D^{\ast
}}\left( \overline{m}^{2}-\frac{m_{K}^{2}}{2}\right) \frac{e^{-\overline{m}%
^{2}/M^{2}}}{M^{2}}+\cdots .  \label{eq:Borel1}
\end{equation}%
The ellipses in Eq.\ (\ref{eq:Borel1}) stand not only for terms suppressed
after this operation, but also for contributions which remain unsuppressed
even after Borel transformation. Therefore, before performing usual
subtraction procedure, one should remove these contributions from $\mathcal{B%
}\Pi ^{\mathrm{Phys}}(p^{2})$. To this end, we have to apply the operator
\begin{equation}
\mathcal{P}(M^{2},m^{2})=\left( 1-M^{2}\frac{d}{dM^{2}}\right)
M^{2}e^{m^{2}/M^{2}},  \label{eq:Operator}
\end{equation}%
to both sides of a sum rule equality \cite{Belyaev:1994zk,Ioffe:1983ju}, and
subtract conventional terms in a usual way.

Then the sum rule for the strong coupling $g_{1}$ reads
\begin{equation}
g_{1}=\frac{2}{fmf_{D^{\ast }}m_{D^{\ast }}(2\overline{m}^{2}-m_{K}^{2})}%
\mathcal{P}(M^{2},\overline{m}^{2})\Pi ^{\mathrm{OPE}}(M^{2},s_{0}),
\label{eq:SC1}
\end{equation}%
where $\Pi ^{\mathrm{OPE}}(M^{2},s_{0})$ is the Borel transformed and
subtracted invariant amplitude $\Pi ^{\mathrm{OPE}}(p^{2})$ that corresponds
to the structure $g_{\mu \nu }$ in $\Pi _{\mu \nu }^{\mathrm{OPE}}(p,q).$

To finish calculation of the strong coupling $g_{1}$, we need to specify
local matrix elements of the $K$ meson which contribute to the function $\Pi
_{\mu \nu }^{\mathrm{OPE}}(p,q)$. Details of calculations necessary to find $%
\Pi _{\mu \nu }^{\mathrm{OPE}}(p,q)$ in the soft limit were presented in
Refs.\ \cite{Agaev:2016dev}, therefore we skip further features of relevant
analysis and provide only final expressions. First of all, our computations
demonstrate that in the soft-meson approximation the correlator $\Pi _{\mu
\nu }^{\mathrm{OPE}}(p,q=0)$ receives contribution from the matrix element
\begin{equation}
\langle 0|\overline{u}i\gamma _{5}s|K\rangle =\frac{f_{K}m_{K}^{2}}{m_{s}},
\label{eq:MatElK1}
\end{equation}%
with $m_{K}$ and $f_{K}$ being the mass and decay constant of the $K^{+}$
meson.

The Borel transformed and subtracted correlation function $\Pi ^{\mathrm{OPE}%
}(M^{2},s_{0})$ is calculated by taking into account condensates up to
dimension $9$ and given below
\begin{eqnarray}
\Pi ^{\mathrm{OPE}}(M^{2},s_{0}) &=&\frac{\mu _{K}}{48\pi ^{2}}\int_{%
\mathcal{M}^{2}}^{s_{0}}\frac{ds(m_{c}^{2}-s)^{2}(m_{c}^{2}+2s)}{s^{2}}%
e^{-s/M^{2}}  \notag \\
&&+\mu _{K}m_{c}\mathcal{F}^{\mathrm{non-pert.}}(M^{2}),  \label{eq:DecayCF}
\end{eqnarray}%
where $\mu _{K}=f_{K}m_{K}^{2}/m_{s}$. In expression above, the
nonperturbative function $\mathcal{F}^{\mathrm{non-pert.}}(M^{2})$ is
determined by the formula%
\begin{eqnarray}
&&\mathcal{F}^{\mathrm{non-pert.}}(M^{2})=-\frac{\langle \overline{d}%
d\rangle }{6}e^{-m_{c}^{2}/M^{2}}+\frac{\langle \frac{\alpha _{s}G^{2}}{\pi }%
\rangle m_{c}}{144M^{4}}  \notag \\
&&\times \int_{0}^{1}\frac{dx\left[ m_{c}^{2}+M^{2}x(1-x)\right] }{%
x^{3}(1-x)^{3}}e^{-m_{c}^{2}/[M^{2}x(1-x)]}  \notag \\
&&-\frac{\langle \overline{d}g\sigma Gd\rangle m_{c}^{2}}{24M^{4}}%
e^{-m_{c}^{2}/M^{2}}+\langle \frac{\alpha _{s}G^{2}}{\pi }\rangle \langle
\overline{d}d\rangle  \notag \\
&&\times \frac{(m_{c}^{2}+3M^{2})\pi ^{2}}{108M^{6}}e^{-m_{c}^{2}/M^{2}}-%
\langle \frac{\alpha _{s}G^{2}}{\pi }\rangle \langle \overline{d}g\sigma
Gd\rangle  \notag \\
&&\times \frac{(5m_{c}^{4}+24m_{c}^{2}M^{2}+6M^{4})\pi ^{2}}{432M^{10}}%
e^{-m_{c}^{2}/M^{2}}.  \label{eq:DecayNPCF}
\end{eqnarray}

\begin{table}[tbp]
\begin{tabular}{|c|c|}
\hline\hline
Quantity & Value (in $\mathrm{MeV}$ units) \\ \hline
$m_{D^{\ast}}=m[D^{\ast}(2010)^{-}]$ & $2010.26\pm 0.05$ \\
$m_{K}=m[K^{+}]$ & $493.677\pm 0.016$ \\
$m_{1}=m[\overline{D}^{\ast}(2007)^{0}]$ & $2006.85\pm 0.05$ \\
$m_{2}=m[K^{0}]$ & $497.611\pm 0.013$ \\
$m_{3}=m[D^{-}]$ & $1869.66\pm 0.05$ \\
$m_{4}=m[K^{\ast}(892)^{+}]$ & $891.67\pm 0.26$ \\
$m_{5}=m[\overline{D}^{0}]$ & $1864.84\pm 0.05$ \\
$m_{6}=m[K^{\ast}(892)^{0}]$ & $895.5\pm 0.8$ \\
$m_{7}=m[\overline{D}_{0}^{\ast}(2300)^{0}]$ & $2343\pm 10$ \\
$f_{K}$ & $155.7 \pm 0.3$ \\
$f_{K^{\ast}}$ & $204 \pm 0.3$ \\
$f_{D}$ & $212.6 \pm 0.7$ \\
$f_{D^{\ast}}$ & $263 \pm 21$ \\
$f_{D^{\ast}_{0}}$ & $373 \pm 19$ \\ \hline\hline
\end{tabular}%
\caption{Masses and decay constants of the mesons $D$, $D^{*}$, $D^{*}_{0}$,
$K$, and $K^{*}$, which have been used in numerical computations.}
\label{tab:Param}
\end{table}

Partial width of the process $X_{\mathrm{AV}}\rightarrow D^{\ast
}(2010)^{-}K^{+}$ can be obtained by employing the following expression
\begin{equation}
\Gamma \left[ X_{\mathrm{AV}}\rightarrow D^{\ast }(2010)^{-}K^{+}\right] =%
\frac{g_{1}^{2}m_{D^{\ast }}^{2}\lambda }{24\pi }\left( 3+\frac{2\lambda ^{2}%
}{m_{D^{\ast }}^{2}}\right),  \label{eq:PartDW}
\end{equation}%
where $\lambda =\lambda \left( m,m_{D^{\ast }},m_{K}\right) $ and
\begin{eqnarray}
\lambda \left( a,b,c\right) &=&\frac{1}{2a}\left[ a^{4}+b^{4}+c^{4}\right.
\notag \\
&&\left. -2\left( a^{2}b^{2}+a^{2}c^{2}+b^{2}c^{2}\right) \right] ^{1/2}.
\end{eqnarray}

The sum rule for the strong coupling $g_{1}$ contains different vacuum
condensates, numerical values of which have been collected in Eq.\ (\ref%
{eq:Parameters}). Apart from that, the equation\ (\ref{eq:SC1}) depends on
the spectroscopic parameters of particles involved into decay process. The
mass and current coupling of the tetraquark $X_{\mathrm{AV}}$ have been
calculated in the current work. The masses and decay constants of the mesons
$D^{\ast }(2010)^{-}$ and $K^{+}$ are collected in Table \ref{tab:Param}.
This table contains also spectroscopic parameters of other mesons which
appear at final stages of different decay channels. For masses all of the
mesons and decay constants of $K$, $K^{\ast }$ and $D$ mesons, we use
information from Ref.\ \cite{PDG:2020}. The decay constant $f_{D^{\ast }}$
of the vector mesons $D^{\ast }(2010)^{-}$ and $D^{\ast }(2007)^{0}$ , and
decay constant $f_{D_{0}^{\ast }}$ of the scalar meson $\overline{D}%
_{0}^{\ast }(2300)^{0}$ are borrowed from Ref.\ \cite{Wang:2015mxa}.

The Borel and continuum subtraction parameters $M^{2}$ and $s_{0}$ required
for calculation of the coupling $g_{1}$ are chosen in accordance with Eq.\ (%
\ref{eq:Wind1}). Numerical computations yield
\begin{equation}
g_{1}=(3.45\pm 0.52)\times 10^{-1}~\mathrm{GeV}^{-1},  \label{eq:Coupl1}
\end{equation}%
Then it is not difficult to find
\begin{equation}
\Gamma \left[ X_{\mathrm{AV}}\rightarrow D^{\ast }(2010)^{-}K^{+}\right]
=(10.6\pm 3.4)~\mathrm{MeV}.  \label{eq:DW1}
\end{equation}

The decay of the tetraquark $X_{\mathrm{AV}}$ to a meson pair $\overline{D}%
^{\ast }(2007)^{0}K^{0}$ is another process with $K$ meson in the final
state. This process is a "neutral" version of the first channel differences
being encoded in masses $m_{1}$ and $m_{2}$ of mesons $\overline{D}^{\ast
}(2007)^{0}$ and $K^{0}$, respectively. Treatment of this decay mode does
not differ from analysis described above. Therefore, we provide final
results for the coupling $g_{2}$
\begin{equation}
g_{2}=(3.67\pm 0.55)\times 10^{-1}~\mathrm{GeV}^{-1},  \label{eq:Coupl2}
\end{equation}%
and partial width of the process
\begin{equation}
\Gamma \left[ X_{\mathrm{AV}}\rightarrow \overline{D}^{\ast }(2007)^{0}K^{0}%
\right] =(11.9\pm 3.8)~\mathrm{MeV}.  \label{eq:DW2}
\end{equation}

\begin{table}[tbp]
\begin{tabular}{|c|c|c|c|}
\hline\hline
$i$ & Channels & $g_{i}~(\mathrm{GeV}^{-1})$ & $\Gamma_{\mathrm{AV}}^{i}~(%
\mathrm{MeV})$ \\ \hline
$1$ & $X_{\mathrm{AV}}\to D^{\ast }(2010)^{-}K^{+}$ & $(3.45 \pm 0.52)\times
10^{-1}$ & $10.6 \pm 3.4$ \\
$2$ & $X_{\mathrm{AV}}\to \overline{D}^{\ast }(2007)^{0}K^{0}$ & $(3.67 \pm
0.55)\times 10^{-1}$ & $11.9 \pm 3.8$ \\
$3$ & $X_{\mathrm{AV}}\to D^{-}K^{\ast}(892)^{+}$ & $1.52 \pm 0.23$ & $17.1
\pm 5.7 $ \\
$4$ & $X_{\mathrm{AV}}\to \overline{D}^{0}K^{\ast }(892)^{0}$ & $1.55 \pm
0.25$ & $18.1 \pm 6.2$ \\ \hline\hline
$j$ &  & $G_{j}$ & $\Gamma_{\mathrm{PS}}^{j}~(\mathrm{MeV})$ \\ \hline
$1$ & $X_{\mathrm{PS}}\to \overline{D}_{0}^{\ast }(2300)^{0}K^{0}$ & $(4.41
\pm 0.66)\times 10^{-1 \star}$ & $16.6 \pm 5.3$ \\
$2$ & $X_{\mathrm{PS}}\to D^{-}K^{\ast }(892)^{+}$ & $1.67 \pm 0.35$ & $23.6
\pm 7.6$ \\
$3$ & $X_{\mathrm{PS}}\to \overline{D}^{0}K^{\ast}(892)^{0}$ & $1.70 \pm
0.38 $ & $24.5 \pm 7.9$ \\ \hline\hline
\end{tabular}%
\caption{Decay channels of the tetraquarks $X_{\mathrm{AV}}$ and $X_{\mathrm{%
PS}}$, strong couplings $g_{i}$, $G_{j}$ and partial widths $\Gamma_{\mathrm{%
AV}}^{i}$, $\Gamma_{\mathrm{PS}}^{j}$. The star-marked coupling $G_1$ has a
dimension $\mathrm{GeV}^{-1}$.}
\label{tab:Channels}
\end{table}

The remaining two decay channels $X_{\mathrm{AV}}\rightarrow D^{-}K^{\ast
}(892)^{+}$ and $X_{\mathrm{AV}}\rightarrow \overline{D}^{0}K^{\ast
}(892)^{0}$ have been explored by a similar manner. Let us consider, for
instance, the process $X_{\mathrm{AV}}\rightarrow D^{-}K^{\ast }(892)^{+}$.
The strong coupling $g_{3}$ that corresponds to the vertex $X_{\mathrm{AV}%
}D^{-}K^{\ast }(892)^{+}$ is defined by the matrix element
\begin{eqnarray}
\langle D^{-}\left( p\right) K^{\ast }(q)|X_{\mathrm{AV}}(p^{\prime
})\rangle &=&g_{3}\left[ \left( q\cdot p^{\prime }\right) \left( \epsilon
^{\ast }\cdot \varepsilon ^{\prime }\right) \right.  \notag \\
&&\left. -\left( q\cdot \varepsilon ^{\prime }\right) \left( p^{\prime
}\cdot \epsilon ^{\ast }\right) \right] ,  \label{eq:Mel3}
\end{eqnarray}%
where $\epsilon _{\mu }^{\ast }$ is polarization vector of the meson $%
K^{\ast }(892)^{+}$. The correlation function that allows us to extract $%
g_{3}$ is

\begin{equation}
\Pi _{\nu }(p,q)=i\int d^{4}xe^{ipx}\langle K^{\ast }(q)|\mathcal{T}%
\{J^{D}(x)J_{\nu }^{\dag }(0)\}|0\rangle ,  \label{eq:CorrF7}
\end{equation}%
with $J^{D}(x)$ being the interpolating current for the pseudoscalar meson $%
D^{-}$%
\begin{equation}
J^{D}(x)=\overline{c}_{j}(x)i\gamma _{5}d_{j}(x).
\end{equation}%
The main term that contributes to this correlation function and determines
phenomenological side of a sum rule for $g_{3}$ has the following form%
\begin{eqnarray}
&&\Pi _{\nu }^{\mathrm{Phys}}(p,q)=g_{3}\frac{fmf_{D}m_{3}^{2}}{m_{c}\left(
p^{2}-m_{3}^{2}\right) \left( p^{\prime 2}-m^{2}\right) }  \notag \\
&&\times \left( \frac{m^{2}+m_{3}^{2}-m_{4}^{2}}{2}\epsilon _{\nu }^{\ast
}+p^{\prime }\cdot \epsilon ^{\ast }q_{\nu }\right) +\cdots ,
\end{eqnarray}%
where $m_{3}$ and $m_{4}$ are masses of $D^{-}$ and $K^{\ast }(892)^{+}$,
respectively. Here, $f_{D}$ is the decay constant of the meson $D^{-}$. To
find $\Pi _{\nu }^{\mathrm{Phys}}(p,q)$, we use the matrix elements of the
tetraquark $X_{\mathrm{AV}}$ and vertex, as well as new matrix element
\begin{equation}
\langle 0|J^{D}(x)|D^{-}\rangle =\frac{f_{D}m_{3}^{2}}{m_{c}}.
\label{eq:DMel}
\end{equation}

The same function $\Pi _{\nu }(p,q)$ calculated in term of quark-gluon
degrees of freedom gives QCD side of the sum rule
\begin{eqnarray}
&&\Pi _{\nu }^{\mathrm{OPE}}(p,q)=-i^{2}\int d^{4}xe^{ipx}\epsilon
\widetilde{\epsilon }\left[ \gamma _{5}\widetilde{S}_{d}^{jc}(x){}\gamma
_{5}\right.  \notag \\
&&\left. \times \widetilde{S}_{c}^{dj}(-x){}\gamma _{\nu }\right] _{\alpha
\beta }\langle K^{\ast }(q)|\overline{u}_{\alpha }^{b}(0)s_{\beta
}^{e}(0)|0\rangle .
\end{eqnarray}%
Our analysis demonstrates that in the soft-meson approximation a
contribution to $\Pi _{\nu }^{\mathrm{OPE}}(p,q)$ comes from the local
matrix element of the meson $K^{\ast }(892)^{+}$%
\begin{equation}
\langle 0|\overline{s}(0)\gamma _{\mu }u(0)|K^{\ast }\rangle =\epsilon _{\mu
}f_{K^{\ast }}m_{4},  \label{eq:KstMel}
\end{equation}%
with $f_{K^{\ast }}$ being its decay constant.

The sum rule for $g_{3}$ can be determined using structures proportional to $%
\epsilon _{\nu }^{\ast }$ in $\Pi _{\nu }^{\mathrm{Phys}}(p,q)$ and $\Pi
_{\nu }^{\mathrm{OPE}}(p,q)$, and corresponding invariant amplitudes $%
\widetilde{\Pi }^{\mathrm{Phys}}(p^{2})$ and $\widetilde{\Pi }^{\mathrm{OPE}%
}(p^{2})$. \ The amplitude $\widetilde{\Pi }^{\mathrm{OPE}}(p^{2})$ is
calculated by including effects of condensates up to dimension $9$. After
Borel transformation and subtraction it takes the form
\begin{eqnarray}
\widetilde{\Pi }^{\mathrm{OPE}}(M^{2},s_{0}) &=&\frac{f_{K^{\ast }}m_{4}}{%
16\pi ^{2}}\int_{\mathcal{M}^{2}}^{s_{0}}\frac{ds(m_{c}^{2}-s)^{2}}{s}%
e^{-s/M^{2}}  \notag \\
&&+m_{4}m_{c}f_{K^{\ast }}\widetilde{\mathcal{F}}^{\mathrm{non-pert.}%
}(M^{2}).  \label{eq:CorrF7a}
\end{eqnarray}%
The function $\widetilde{\mathcal{F}}^{\mathrm{non-pert.}}(M^{2})$ in Eq.\ (%
\ref{eq:CorrF7a}) is given by the expression%
\begin{eqnarray}
&&\widetilde{\mathcal{F}}^{\mathrm{non-pert.}}(M^{2})=-\frac{\langle
\overline{d}d\rangle }{6}e^{-m_{c}^{2}/M^{2}}+\frac{\langle \frac{\alpha
_{s}G^{2}}{\pi }\rangle m_{c}^{3}}{144M^{4}}  \notag \\
&&\times \int_{0}^{1}\frac{dxe^{-m_{c}^{2}/[M^{2}x(1-x)]}}{x^{3}(x-1)^{3}}+%
\frac{\langle \overline{d}g\sigma Gd\rangle m_{c}^{2}}{24M^{4}}%
e^{-m_{c}^{2}/M^{2}}  \notag \\
&&-\langle \frac{\alpha _{s}G^{2}}{\pi }\rangle \langle \overline{d}d\rangle
\frac{(m_{c}^{2}+3M^{2})\pi ^{2}}{108M^{6}}e^{-m_{c}^{2}/M^{2}}+\langle
\frac{\alpha _{s}G^{2}}{\pi }\rangle  \notag \\
&& \times \langle \overline{d}g\sigma Gd\rangle \frac{%
(5m_{c}^{4}+24m_{c}^{2}M^{2}+6M^{4})\pi ^{2}}{432M^{10}}e^{-m_{c}^{2}/M^{2}}.
\notag \\
&&
\end{eqnarray}%
After manipulations described above in detail, for $g_{3}$ we get the sum
rule%
\begin{equation}
g_{3}=\frac{2m_{c}}{fmf_{D}m_{3}^{2}(2\widetilde{m}^{2}-m_{K^{\ast }}^{2})}%
\mathcal{P}(M^{2},\widetilde{m}^{2})\widetilde{\Pi }^{\mathrm{OPE}%
}(M^{2},s_{0}),  \label{eq:SC2}
\end{equation}%
where $\widetilde{m}^{2}=(m^{2}+m_{3}^{2})/2$.

The sum rule prediction for $g_{3}$ reads
\begin{equation}
g_{3}=(1.52\pm 0.23)~\mathrm{GeV}^{-1},  \label{eq:Coupl3}
\end{equation}%
The partial width of the process $X_{\mathrm{AV}}\rightarrow D^{-}K^{\ast
}(892)^{+}$ can be computed using Eq.\ (\ref{eq:PartDW}) after replacing $%
g_{1}$,$\lambda $, and $m_{D^{\ast }}$ by $g_{3}$, $\widetilde{\lambda }$,
and $m_{4}$, where $\widetilde{\lambda }=\lambda (m^{2},m_{3}^{2},m_{4}^{2})$%
. Calculations lead to the following result
\begin{equation}
\Gamma \left[ X_{\mathrm{AV}}\rightarrow D^{-}K^{\ast }(892)^{+}\right]
=(17.1\pm 5.7)~\mathrm{MeV}.  \label{eq:DW3}
\end{equation}%
Predictions for the strong coupling $g_{4}$ and partial with of the decay $%
X_{\mathrm{AV}}\rightarrow \overline{D}^{0}K^{\ast }(892)^{0}$, as well as
results obtained in the present section are collected in Table\ \ref%
{tab:Channels}. This information allows us to estimate full width of the
tetraquark $X_{\mathrm{AV}}$
\begin{equation}
\Gamma _{\mathrm{AV}}=(58\pm 10)~\mathrm{MeV},  \label{eq:FWAV}
\end{equation}%
which characterizes it as a resonance with a relatively narrow width.

%%%%%%%%%%%%%%%%%%%%%%%%%%%%%%%%%%%%%%%%%%%%%%%%%%%%%%%%%%%%%%%%%%%%%%%%%%%%

\section{Processes $X_{\mathrm{PS}}\rightarrow \overline{D}_{0}^{\ast
}(2300)^{0}K^{0}$, $D^{-}K^{\ast }(892)^{+}$ and $\overline{D}^{0}K^{\ast
}(892)^{0}$}

\label{sec:Decays2}
%%%%%%%%%%%%%%%%%%%%%%%%%%%%%%%%%%%%%%%%%%%%%%%%%%%%%%%%%%%

In this section, we investigate decays of the pseudoscalar tetraquark $X_{%
\mathrm{PS}}$ with the mass $m_{\mathrm{PS}}$ and current coupling $f_{%
\mathrm{PS}}$ which have been extracted from two-point sum rules in Section %
\ref{sec:Masses}. We consider the $S$-wave process $X_{\mathrm{PS}%
}\rightarrow \overline{D}_{0}^{\ast }(2300)^{0}K^{0}$, as well as $P$-wave
decay modes $X_{\mathrm{PS}}\rightarrow D^{-}K^{\ast }(892)^{+}$ and $X_{%
\mathrm{PS}}\rightarrow \overline{D}^{0}K^{\ast }(892)^{0}$ of this
four-quark state.

We begin our analysis from the decay $X_{\mathrm{PS}}\rightarrow \overline{D}%
_{0}^{\ast }(2300)^{0}K^{0}$. The coupling $G_{1}$ required to calculate
partial width of this process can be defined using the on-mass-shell matrix
element
\begin{equation}
\langle D_{0}^{\ast }\left( p\right) K^{0}(q)|X_{\mathrm{PS}}(p^{\prime
})\rangle =G_{1}p\cdot p^{\prime }.  \label{eq:Mel4}
\end{equation}%
The correlation function, which should be considered to determine strong
coupling $G_{1}$ of particles at the vertex $X_{\mathrm{PS}}\overline{D}%
_{0}^{\ast }(2300)^{0}K^{0}$, is given by the formula
\begin{equation}
\widehat{\Pi }(p,q)=i\int d^{4}xe^{ipx}\langle K^{0}(q)|\mathcal{T}%
\{J^{D_{0}^{\ast }}(x)J_{\mathrm{PS}}^{\dag }(0)\}|0\rangle ,
\label{eq:CorrF8}
\end{equation}%
where $D_{0}^{\ast }$ stands for meson $\overline{D}_{0}^{\ast }(2300)^{0}$.
Here, $J_{\mathrm{PS}}(x)$ and $J^{D_{0}^{\ast }}(x)$ are interpolating
currents for the tetraquark $X_{\mathrm{PS}}$ [see, Eq.\ (\ref{eq:CR2})],
and for the scalar meson $\overline{D}_{0}^{\ast }(2300)^{0}$. The latter is
defined by the expression
\begin{equation}
J^{D_{0}^{\ast }}(x)=\overline{c}_{j}(x)d_{j}(x).
\end{equation}%
A contribution to the correlation function $\widehat{\Pi }^{\mathrm{Phys}%
}(p,q)$ with poles at $p^{2}$ and $p^{\prime 2}=(p+q)^{2}$ comes from the
term
\begin{eqnarray}
\widehat{\Pi }^{\mathrm{Phys}}(p,q) &=&G_{1}\frac{f_{\mathrm{PS}}m_{\mathrm{%
PS}}^{2}f_{D_{0}^{\ast }}m_{7}}{(m_{c}+m_{s})\left( p^{2}-m_{7}^{2}\right)
\left( p^{\prime 2}-m_{\mathrm{PS}}^{2}\right) }  \notag \\
&&\times \frac{m_{\mathrm{PS}}^{2}+m_{7}^{2}-m_{2}^{2}}{2}+\cdots .
\label{eq:CorrF9}
\end{eqnarray}%
Here, $m_{7}$ and $f_{D_{0}^{\ast }}$ are the mass and decay constant of the
meson $\overline{D}_{0}^{\ast }(2300)^{0}$, respectively. In order to find
Eq.\ (\ref{eq:CorrF9}), we employ Eqs.\ (\ref{eq:Mel4}) and (\ref{eq:Mel1a}%
), as well as the matrix element
\begin{equation}
\langle 0|J^{D_{0}^{\ast }}|D_{0}^{\ast }(p)\rangle =f_{D_{0}^{\ast }}m_{7}.
\end{equation}

The QCD side of the sum rule $\widehat{\Pi }^{\mathrm{OPE}}(p,q)$ has the
following form%
\begin{eqnarray}
&&\widehat{\Pi }^{\mathrm{OPE}}(p,q)=i\int d^{4}xe^{ipx}\epsilon \widetilde{%
\epsilon }\left[ \widetilde{S}_{u}^{jb}(x){}\widetilde{S}_{c}^{dj}(-x)\gamma
_{5}\right] {}_{\alpha \beta }  \notag \\
&&\times \langle K^{0}(q)|\overline{d}_{\alpha }^{c}(0)s_{\beta
}^{e}(0)|0\rangle .
\end{eqnarray}%
The functions $\widehat{\Pi }^{\mathrm{Phys}}(p,q)$ and $\widehat{\Pi }^{%
\mathrm{OPE}}(p,q)$ have trivial Lorentz structures $\sim I$, therefore both
of them contain only one invariant amplitude. The amplitude $\widehat{\Pi }^{%
\mathrm{OPE}}(p^{2})$ is calculated with dimension-$9$ accuracy and given by
the following expression
\begin{eqnarray}
\widehat{\Pi }^{\mathrm{OPE}}(M^{2},s_{0}) &=&\frac{\mu _{K^{0}}}{16\pi ^{2}}%
\int_{\mathcal{M}^{2}}^{s_{0}}\frac{ds(m_{c}^{2}-s)^{2}}{s}e^{-s/M^{2}}
\notag \\
&&+\mu _{K^{0}}m_{c}\widehat{\mathcal{F}}^{\mathrm{non-pert.}}(M^{2}),
\end{eqnarray}%
where the function $\widehat{\mathcal{F}}^{\mathrm{non-pert.}}(M^{2})$ is
determined by the formula
\begin{eqnarray}
&&\widehat{\mathcal{F}}^{\mathrm{non-pert.}}(M^{2})=\frac{\langle \overline{u%
}u\rangle }{16}e^{-m_{c}^{2}/M^{2}}-\frac{\langle \frac{\alpha _{s}G^{2}}{%
\pi }\rangle m_{c}^{3}}{144M^{4}}  \notag \\
&&\times \int_{0}^{1}\frac{dxe^{-m_{c}^{2}/[M^{2}x(1-x)]}}{x^{3}(x-1)^{3}}-%
\frac{\langle \overline{u}g\sigma Gu\rangle m_{c}^{2}}{24M^{4}}%
e^{-m_{c}^{2}/M^{2}}  \notag \\
&&+\langle \frac{\alpha _{s}G^{2}}{\pi }\rangle \langle \overline{u}u\rangle
\frac{(m_{c}^{2}+3M^{2})\pi ^{2}}{108M^{6}}e^{-m_{c}^{2}/M^{2}}-\langle
\frac{\alpha _{s}G^{2}}{\pi }\rangle  \notag \\
&&\times \langle \overline{u}g\sigma Gu\rangle \frac{%
(5m_{c}^{4}+24m_{c}^{2}M^{2}+6M^{4})\pi ^{2}}{432M^{10}}e^{-m_{c}^{2}/M^{2}},
\notag \\
&&  \label{eq:CorrF10}
\end{eqnarray}%
and $\mu _{K^{0}}=m_{2}^{2}f_{K}/m_{s}$.

In the soft-meson approximation the coupling $G_{1}$ can be found by means
of the sum rule
\begin{eqnarray}
G_{1} &=&\frac{2(m_{c}+m_{s})}{f_{\mathrm{PS}}m_{\mathrm{PS}%
}^{2}f_{D_{0}^{\ast }}m_{7}(2\widehat{m}^{2}-m_{K}^{2})}  \notag \\
&&\times \mathcal{P}(M^{2},\widehat{m}^{2})\widehat{\Pi }^{\mathrm{OPE}%
}(M^{2},s_{0}),
\end{eqnarray}%
where $\widehat{m}^{2}=(m_{\mathrm{PS}}^{2}+m_{7}^{2})/2$.

The width of the decay $X_{\mathrm{PS}}\rightarrow \overline{D}_{0}^{\ast
}(2300)^{0}K^{0}$ is found by utilizing the formula%
\begin{equation}
\Gamma \left[ X_{\mathrm{PS}}\rightarrow \overline{D}_{0}^{\ast
}(2300)^{0}K^{0}\right] =G_{1}^{2}\frac{m_{7}^{2}\widehat{\lambda }}{8\pi }%
\left( 1+\frac{\widehat{\lambda }^{2}}{m_{7}^{2}}\right) ,  \label{eq:DW4}
\end{equation}%
in which $\widehat{\lambda }=\lambda (m_{\mathrm{PS}},m_{7},m_{2})$. Our
computations for the coupling $G_{1}$ and partial width of the process yield
\begin{equation}
G_{1}=(4.41\pm 0.66)\times 10^{-1}~\mathrm{GeV}^{-1},
\end{equation}%
and%
\begin{equation}
\Gamma \left[ X_{\mathrm{PS}}\rightarrow \overline{D}_{0}^{\ast
}(2300)^{0}K^{0}\right] =(16.6\pm 5.3)~\mathrm{MeV}.  \label{eq:DW5}
\end{equation}

The coupling $G_{2}$ that describes strong interaction of particles at the
vertex $X_{\mathrm{PS}}D^{-}K^{\ast }(892)^{+}$ and determines partial width
of the channel $X_{\mathrm{PS}}\rightarrow D^{-}K^{\ast }(892)^{+}$ is
defined by the matrix element
\begin{equation}
\langle D^{-}\left( p\right) K^{\ast }(q)|X_{\mathrm{PS}}(p^{\prime
})\rangle =G_{2}p\cdot \epsilon ^{\ast }.  \label{eq:Mel5}
\end{equation}%
The sum rule for $G_{2}$ is obtained from the correlation function
\begin{equation}
\Pi ^{\prime }(p,q)=i\int d^{4}xe^{ipx}\langle K^{\ast }(q)|\mathcal{T}%
\{J^{D}(x)J_{\mathrm{PS}}^{\dag }(0)\}|0\rangle .  \label{eq:CorrF11}
\end{equation}%
In terms of involved particles' physical parameters this function has the
form
\begin{eqnarray}
\Pi ^{\prime \mathrm{Phys}}(p,q) &=&G_{2}\frac{f_{\mathrm{PS}}m_{\mathrm{PS}%
}^{2}}{(m_{c}+m_{s})\left( p^{\prime 2}-m_{\mathrm{PS}}^{2}\right) }  \notag
\\
&&\times \frac{f_{D}m_{3}^{2}}{m_{c}\left( p^{2}-m_{3}^{2}\right) }p\cdot
\epsilon ^{\ast }+\cdots .  \label{eq:CorrF12}
\end{eqnarray}%
To extract this expression, we use the matrix elements from Eqs.\ (\ref%
{eq:DMel}) and (\ref{eq:KstMel}), as well as one defined by Eq.\ (\ref%
{eq:Mel5}).

The correlation function $\Pi ^{\prime }(p,q)$ calculated using quark-gluon
degrees of freedom fixes the QCD side of the sum rule and is equal to
\begin{eqnarray}
&&\Pi ^{\prime \mathrm{OPE}}(p,q)=-\int d^{4}xe^{ipx}\epsilon \widetilde{%
\epsilon }\left[ \widetilde{S}_{d}^{jc}(x){}\gamma _{5}\widetilde{S}%
_{c}^{dj}(-x)\gamma _{5}\right] _{\alpha \beta }{}  \notag \\
&&\times \langle K^{\ast }(q)|\overline{u}_{\alpha }^{b}(0)s_{\beta
}^{e}(0)|0\rangle .  \label{eq:CorrF13}
\end{eqnarray}%
The sum rule for the strong coupling $G_{2}$ can be obtained by employing
standard manipulations. The width of the process $X_{\mathrm{PS}}\rightarrow
D^{-}K^{\ast }(892)^{+}$ is calculated by means of the expression%
\begin{equation}
\Gamma \left[ X_{\mathrm{PS}}\rightarrow D^{-}K^{\ast }(892)^{+}\right]
=G_{2}^{2}\frac{\lambda ^{\prime 3}}{8\pi m_{4}^{2}},  \label{eq:DW6}
\end{equation}%
where $\lambda ^{\prime }=\lambda (m_{\mathrm{PS}},m_{3},m_{4})$.

For the coupling $G_{2}$ numerical computations give%
\begin{equation}
G_{2}=1.67\pm 0.35,
\end{equation}%
and the partial width of the decay under analysis equals to
\begin{equation}
\Gamma \left[ X_{\mathrm{PS}}\rightarrow D^{-}K^{\ast }(892)^{+}\right]
=(23.6\pm 7.6)~\mathrm{MeV}.  \label{eq:DW7}
\end{equation}%
The decay channel $X_{\mathrm{PS}}\rightarrow \overline{D}^{0}K^{\ast
}(892)^{0}$ of the tetraquark $X_{\mathrm{PS}}$ which is last process
considered in the present paper, can be treated in a similar way. Therefore,
we refrain from further details and provide all relevant information in
Table\ \ref{tab:Channels}.

The full width of $X_{\mathrm{PS}}$%
\begin{equation}
\Gamma _{\mathrm{PS}}=(65\pm 12)~\mathrm{MeV},  \label{eq:FWPS}
\end{equation}%
does not differ considerably from the width of the axial-vector tetraquark $%
X_{\mathrm{AV}}$.

%%%%%%%%%%%%%%%%%%%%%%%%%%%%%%%%%%%%%%%%%%%%%%%%%%%%%%%%%%%%%%%%%%%%%%%%%%%%

\section{Conclusions}

\label{sec:Disc}
%%%%%%%%%%%%%%%%%%%%%%%%%%%%%%%%%%%%%%%%%%%%%%%%%%%%%%%%%%%

In the current article, we have investigated the axial-vector and
pseudoscalar tetraquarks $X_{\mathrm{AV}}$ and $X_{\mathrm{PS}}$ from a
family of exotic mesons $[ud][\overline{c}\overline{s}]$ containing four
different quark flavors. We have calculated their masses, and also estimated
full widths of these states using different decay channels.

Interest to fully open-flavor structures renewed recently due to discovery
of resonances $X_{0(1)}(2900)$ made by the LHCb collaboration. One of these
states $X_{1}(2900)$ was studied in Ref.\ \cite{Agaev:2021knl} as a vector
tetraquark $X_{\mathrm{V}}=[ud][\overline{c}\overline{s}]$ . The mass and
width of $X_{\mathrm{V}}$ are close to physical parameters of the resonance $%
X_{1}(2900)$ measured by LHCb, which allowed us to interpret $X_{\mathrm{V}}$
as a candidate to the vector resonance $X_{1}(2900)$.

\begin{figure}[h]
\includegraphics[width=8.8cm]{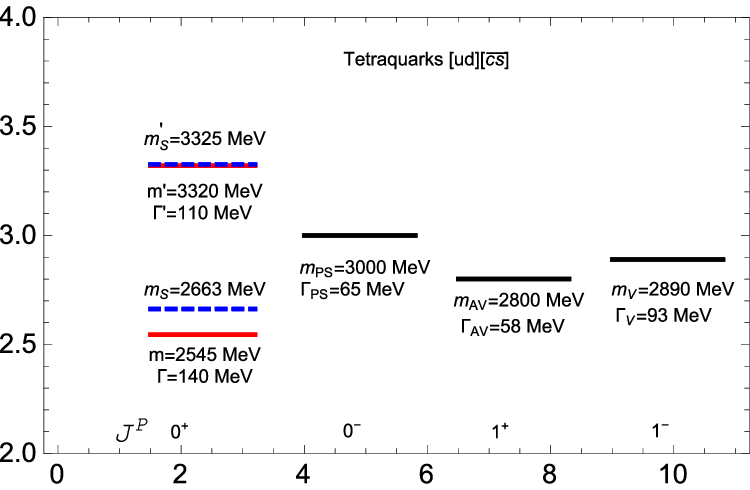}
\caption{Mass and width of tetraquarks $[ud][\overline{c}\overline{s}]$ with
different spin-parities. The lower and upper red (solid) and blue (dashed)
lines are $1S$ and $2S$ scalar states, respectively. The red lines
correspond to the scalar tetraquark $X_{0}$, whereas blue lines show
parameters of $X_{\mathrm{S}}$. The mass and width of the vector particle $%
X_{\mathrm{V}}$ were determined in Ref.\ \protect\cite{Agaev:2021knl}. The
axial-vector and pseudoscalar states have been explored in the current
article. Theoretical uncertainties of extracted observables are not shown.}
\label{fig:Spectr}
\end{figure}
The masses and widths of the ground-state scalar tetraquarks $X_{0}$ and $X_{%
\mathrm{S}}$, and their first radial excitations were calculated in Ref.\
\cite{Agaev:2022eeh}. The scalar particles $X_{0}$ and $X_{\mathrm{S}}$ were
modeled using axial-vector and scalar diquark-antidiquark pairs,
respectively.

Information gained in our studies about tetraquarks $[ud][\overline{c}%
\overline{s}]$ with spin-parities $J^{\mathrm{P}}=0^{+}$, $0^{-}$,$1^{+}$
and $1^{-}$ is shown in Fig.\ \ref{fig:Spectr}. As is seen, the pseudoscalar
$X_{\mathrm{PS}}$ state is heaviest particle in family of tetraquarks $[ud][%
\overline{c}\overline{s}]$, whereas light particles are scalar ones with
different internal organizations. The vector and axial-vector states have
comparable masses and widths.

It is interesting to consider hadronic processes, where exotic mesons $X_{%
\mathrm{AV}}$ and $X_{\mathrm{PS}}$ may be observed. We have noted in
Section \ref{sec:Int} that resonances $X_{0}(2900)$ and $X_{1}(2900)$ were
discovered in the invariant $D^{-}K^{+}$ mass distribution of the process $%
B^{+}\rightarrow D^{+}D^{-}K^{+}$. For the scalar tetraquark $X_{0}$ the
decay $X_{0}\rightarrow D^{-}K^{+}$ is $S$-wave process, whereas $%
X_{1}\rightarrow D^{-}K^{+}$ is $P$-wave channel for the vector particle $%
X_{1}(2900)$. The invariant mass distribution of $D^{-}K^{\ast }(892)^{+}$
mesons in exclusive decays of $B^{+}$ meson may be explored to observe
tetraquarks $X_{\mathrm{AV}}$ and $X_{\mathrm{PS}}$. In fact, decays to $%
D^{-}K^{\ast }(892)^{+}$ mesons are $S$-wave and $P$-wave channels for the
tetraquarks $X_{\mathrm{AV}}$ and $X_{\mathrm{PS}}$, respectively. Because
branching ratios of these channels amount to $0.30$ and $0.36$,
respectively, they may be employed to fix structures $X_{\mathrm{AV}}$ and $%
X_{\mathrm{PS}}$. These and relevant problems require further theoretical
and experimental studies.

\begin{widetext}

%%%%%%%%%%%%%%%%%%%%%%%%%%%%%%%%%%%%%%%%%%%%%%%%%%%%%%%%%%%%%%%%%%%%%%%%
\appendix*

\section{ Quark propagators and invariant amplitude $\Pi (M^{2},s_{0})$}

\renewcommand{\theequation}{\Alph{section}.\arabic{equation}} \label{sec:App}
%%%%%%%%%%%%%%%%%%%%%%%%%%%%%%%%%%%%%%%%%%%%%%%%%%%%%%%%%%%%%%%%%%%%%%

In this work, for the light quark propagator $S_{q}^{ab}(x)$, we use the
expression
\begin{eqnarray}
&&S_{q}^{ab}(x)=i\delta _{ab}\frac{\slashed x}{2\pi ^{2}x^{4}}-\delta _{ab}%
\frac{m_{q}}{4\pi ^{2}x^{2}}-\delta _{ab}\frac{\langle \overline{q}q\rangle
}{12}+i\delta _{ab}\frac{\slashed xm_{q}\langle \overline{q}q\rangle }{48}%
-\delta _{ab}\frac{x^{2}}{192}\langle \overline{q}g_{s}\sigma Gq\rangle
\notag \\
&&+i\delta _{ab}\frac{x^{2}\slashed xm_{q}}{1152}\langle \overline{q}%
g_{s}\sigma Gq\rangle -i\frac{g_{s}G_{ab}^{\alpha \beta }}{32\pi ^{2}x^{2}}%
\left[ \slashed x{\sigma _{\alpha \beta }+\sigma _{\alpha \beta }}\slashed x%
\right] -i\delta _{ab}\frac{x^{2}\slashed xg_{s}^{2}\langle \overline{q}%
q\rangle ^{2}}{7776}  \notag \\
&&-\delta _{ab}\frac{x^{4}\langle \overline{q}q\rangle \langle
g_{s}^{2}G^{2}\rangle }{27648}+\cdots .
\end{eqnarray}%
For the heavy quark $Q=c$, we employ the propagator $S_{Q}^{ab}(x)$
\begin{eqnarray}
&&S_{Q}^{ab}(x)=i\int \frac{d^{4}k}{(2\pi )^{4}}e^{-ikx}\Bigg \{\frac{\delta
_{ab}\left( {\slashed k}+m_{Q}\right) }{k^{2}-m_{Q}^{2}}-\frac{%
g_{s}G_{ab}^{\alpha \beta }}{4}\frac{\sigma _{\alpha \beta }\left( {\slashed %
k}+m_{Q}\right) +\left( {\slashed k}+m_{Q}\right) \sigma _{\alpha \beta }}{%
(k^{2}-m_{Q}^{2})^{2}}  \notag \\
&&+\frac{g_{s}^{2}G^{2}}{12}\delta _{ab}m_{Q}\frac{k^{2}+m_{Q}{\slashed k}}{%
(k^{2}-m_{Q}^{2})^{4}}+\frac{g_{s}^{3}G^{3}}{48}\delta _{ab}\frac{\left( {%
\slashed k}+m_{Q}\right) }{(k^{2}-m_{Q}^{2})^{6}}\left[ {\slashed k}\left(
k^{2}-3m_{Q}^{2}\right) +2m_{Q}\left( 2k^{2}-m_{Q}^{2}\right) \right] \left(
{\slashed k}+m_{Q}\right) +\cdots \Bigg \}.  \notag \\
&&
\end{eqnarray}

Above, we have used the shorthand notations
\begin{equation}
G_{ab}^{\alpha \beta }\equiv G_{A}^{\alpha \beta }\lambda _{ab}^{A}/2,\ \
G^{2}=G_{\alpha \beta }^{A}G_{A}^{\alpha \beta },\ G^{3}=f^{ABC}G_{\alpha
\beta }^{A}G^{B\beta \delta }G_{\delta }^{C\alpha },
\end{equation}%
where $G_{A}^{\alpha \beta }$ is the gluon field strength tensor, $\lambda
^{A}$ and $f^{ABC}$ are the Gell-Mann matrices and structure constants of
the color group $SU_{c}(3)$, respectively. The indices $A,B,C$ run in the
range $1,2,\ldots 8$.

The invariant amplitude $\Pi (M^{2},s_{0})$ obtained after the Borel
transformation and subtraction is equal to%
\begin{equation*}
\Pi (M^{2},s_{0})=\int_{\mathcal{M}^{2}}^{s_{0}}ds\rho ^{\mathrm{OPE}%
}(s)e^{-s/M^{2}}+\Pi (M^{2}),
\end{equation*}%
where the spectral density $\rho ^{\mathrm{OPE}}(s)$ and the function $\Pi
(M^{2})$ are determined by formulas
\begin{equation}
\rho ^{\mathrm{OPE}}(s)=\rho ^{\mathrm{pert.}}(s)+\sum_{N=3}^{8}\rho ^{%
\mathrm{DimN}}(s),\ \ \Pi (M^{2})=\sum_{N=6}^{10}\Pi ^{\mathrm{DimN}}(M^{2}),
\label{eq:A1}
\end{equation}%
respectively. The components of $\rho ^{\mathrm{OPE}}(s)$ and $\Pi (M^{2})$
are given by expressions%
\begin{equation}
\rho ^{\mathrm{DimN}}(s)=\int_{0}^{1}d\alpha \rho ^{\mathrm{DimN}}(s,\alpha
),\ \ \Pi ^{\mathrm{DimN}}(M^{2})=\int_{0}^{1}d\alpha \Pi ^{\mathrm{DimN}%
}(M^{2},\alpha ),  \label{eq:A5}
\end{equation}
where $\alpha $ is the Feynman parameter.

Below, we write down components of the spectral density $\rho ^{\mathrm{OPE}%
}(s)$ and function $\Pi (M^{2})$ for the axial-vector tetraquark $X_{\mathrm{%
AV}}$:

The perturbative and nonperturbative components of the spectral density $%
\rho ^{\mathrm{pert.}}(s,\alpha )$ and $\rho ^{\mathrm{Dim3(4,5,6,7,8)}%
}(s,\alpha )$ have the forms:
\begin{equation}
\rho ^{\mathrm{pert.}}(s,\alpha )=\frac{\Theta (L)}{12288\pi ^{6}(1-\alpha
)^{3}}\left[ m_{c}^{2}-s(1-\alpha )\right] ^{3}\alpha ^{3}\left[
m_{c}^{2}\alpha -16m_{c}m_{s}-5s\alpha (1-\alpha )\right],  \label{eq:A6}
\end{equation}%
\begin{equation}
\rho ^{\mathrm{Dim3}}(s,\alpha )=-\frac{\langle \overline{s}s\rangle \Theta
(L)\alpha ^{2}}{128\pi ^{4}(1-\alpha )^{2}}\left[ m_{c}^{2}-s(1-\alpha )%
\right] \left[ 2m_{c}^{3}-m_{c}^{2}m_{s}(1-\alpha )-2m_{c}s(1-\alpha
)+3m_{s}s(1-\alpha )^{2}\right],  \label{eq:A7}
\end{equation}%
\begin{eqnarray}
&&\rho ^{\mathrm{Dim4}}(s,\alpha )=\frac{\langle \alpha _{s}G^{2}/\pi
\rangle \Theta (L)\alpha }{55296\pi ^{4}(1-\alpha )^{3}}\left[ s^{2}\alpha
(1-\alpha )^{3}(162-167\alpha )+3m_{c}^{4}\alpha (18-37\alpha +21\alpha
^{2})\right.  \notag \\
&&-24m_{c}^{3}m_{s}(9-18\alpha +9\alpha ^{2}+2\alpha ^{3})+4sm_{c}^{2}\alpha
(-54+164\alpha -169\alpha ^{2}+59\alpha ^{3})  \notag \\
&&\left. -24sm_{c}m_{s}(-9+27\alpha -26\alpha ^{2}+5\alpha ^{3}+3\alpha
^{4}) \right],  \label{eq:A8}
\end{eqnarray}%
\begin{equation}
\rho ^{\mathrm{Dim5}}(s,\alpha )=-\frac{\langle \overline{s}g_{s}\sigma
Gs\rangle \Theta (L)\alpha }{192\pi ^{4}(1-\alpha )}\left[
3m_{c}^{3}-m_{c}^{2}m_{s}(1-\alpha )-3sm_{c}(1-\alpha )+2sm_{s}(1-\alpha
)^{2}\right],  \label{eq:A9}
\end{equation}%
\begin{eqnarray}
&&\rho ^{\mathrm{Dim6}}(s,\alpha )=-\frac{\Theta (L)}{405\cdot 2^{12}\pi
^{6}(1-\alpha )^{3}}\left\{ (1-\alpha )^{3}\left[ -m_{c}m_{s}+m_{c}^{2}%
\alpha -2s\alpha (1-\alpha )\right] \right.  \notag \\
&&\times (1280\langle \overline{d}d\rangle ^{2}g_{s}^{2}\pi
^{2}+138240\langle \overline{d}d\rangle \langle \overline{u}u\rangle \pi
^{4})-\alpha \left[ 1280g_{s}^{2}\pi ^{2}\langle \overline{s}s\rangle
^{2}(m_{c}^{2}-2s(1-\alpha ))(\alpha -1)^{3}\right.  \notag \\
&&\left. \left. +27\langle g_{s}^{3}G^{3}\rangle m_{c}^{2}\alpha ^{4}\right]
\right\},  \label{eq:A10}
\end{eqnarray}%
\begin{equation}
\rho ^{\mathrm{Dim7}}(s,\alpha )=-\frac{\langle \alpha _{s}G^{2}/\pi \rangle
\langle \overline{s}s\rangle \Theta (L)}{1152\pi ^{2}(1-\alpha )^{2}}\left[
4m_{c}(2-4\alpha +2\alpha ^{2}+\alpha ^{3})-3m_{s}(1-\alpha )^{3}\right],
\label{eq:A11}
\end{equation}%
\begin{equation}
\rho ^{\mathrm{Dim8}}(s,\alpha )=\frac{\Theta (L)}{18432\pi ^{2}}\left[
768\langle \overline{u}u\rangle \langle \overline{d}g_{s}\sigma Gd\rangle
(\alpha -1)-\langle \alpha _{s}G^{2}/\pi \rangle ^{2}\alpha \right].
\label{eq:A12}
\end{equation}

Components of the function $\Pi (M^{2})$ are:%
\begin{eqnarray}
&&\Pi ^{\mathrm{Dim6}}(M^{2},\alpha )=\frac{\langle g_{s}^{3}G^{3}\rangle
m_{c}^{3}\alpha ^{3}}{184320\pi ^{6}M^{2}(1-\alpha )^{5}}\exp \left[ -\frac{%
m_{c}^{2}}{M^{2}(1-\alpha )}\right] \left[ m_{c}^{3}\alpha (2+\alpha
)+8m_{c}^{2}m_{s}(2\alpha -1)\right.  \notag \\
&&\left. -16M^{2}m_{s}(\alpha ^{2}-1)\right],  \label{eq:A13}
\end{eqnarray}%
\begin{equation}
\Pi ^{\mathrm{Dim7}}(M^{2},\alpha )=-\frac{\langle \alpha _{s}G^{2}/\pi
\rangle \langle \overline{s}s\rangle m_{c}^{3}\alpha ^{2}}{1152M^{2}\pi
^{2}(1-\alpha )^{3}}\exp \left[ -\frac{m_{c}^{2}}{M^{2}(1-\alpha )}\right] %
\left[ m_{c}m_{s}-2M^{2}(1-\alpha )\right],  \label{eq:A14}
\end{equation}%
\begin{eqnarray}
&&\Pi ^{\mathrm{Dim8}}(M^{2},\alpha ) =-\frac{\langle \alpha _{s}G^{2}/\pi
\rangle ^{2}m_{c}\alpha }{9216M^{2}\pi ^{2}(1-\alpha )^{3}}\exp \left[ -%
\frac{m_{c}^{2}}{M^{2}(1-\alpha )}\right] \left[ 2m_{c}^{2}m_{s}(1-\alpha
)+m_{c}^{3}\alpha +2M^{2}m_{s}(\alpha ^{2}-1)\right]  \notag \\
&&-\frac{\langle \overline{u}u\rangle \langle \overline{d}g_{s}\sigma
Gd\rangle m_{c}m_{s}}{24\pi ^{2}}\exp \left[ -\frac{m_{c}^{2}}{M^{2}}\right],
\label{eq:A15}
\end{eqnarray}

\begin{eqnarray}
&&\Pi ^{\mathrm{Dim9}}(M^{2},\alpha )=\frac{m_{c}\alpha }{69120M^{6}\pi
^{4}(1-\alpha )^{5}}\exp \left[ -\frac{m_{c}^{2}}{M^{2}(1-\alpha )}\right]
\left\{ 20\langle \alpha _{s}G^{2}/\pi \rangle \langle \overline{s}%
g_{s}\sigma Gs\rangle M^{2}\pi ^{2}(1-\alpha )^{2}\right.  \notag \\
&&\times \left[ m_{c}^{3}m_{s}-3m_{c}^{2}M^{2}(1-\alpha
)-m_{c}m_{s}M^{2}(1-\alpha )-3M^{4}(\alpha ^{2}-1)\right] +3\langle
g_{s}^{3}G^{3}\rangle \langle \overline{s}s\rangle m_{c}\alpha \left[
4m_{c}^{3}M^{2}(1-2\alpha )\right.  \notag \\
&&\left. \left. -16m_{c}M^{4}(1-\alpha )+m_{c}^{4}m_{s}(2+\alpha
)+m_{c}^{2}m_{s}M^{2}(\alpha ^{2}+\alpha -2)-m_{s}M^{4}(2-3\alpha +\alpha
^{3})\right] \right\}  \notag \\
&&+\frac{m_{c}}{62208\pi ^{2}M^{2}}\exp \left[ -\frac{m_{c}^{2}}{M^{2}}%
\right] \left\{ -27\langle \alpha _{s}G^{2}/\pi \rangle \langle \overline{s}%
g_{s}\sigma Gs\rangle (m_{c}m_{s}-3M^{2})+32g_{s}^{2}\langle \overline{s}%
s\rangle (\langle \overline{u}u\rangle ^{2}+\langle \overline{d}d\rangle
^{2})\right.  \notag \\
&&\left. \times (m_{c}m_{s}-2M^{2})+3456\langle \overline{s}s\rangle \langle
\overline{u}u\rangle \langle \overline{d}d\rangle \pi
^{2}(m_{c}m_{s}-2M^{2})\right\},  \label{eq:A16}
\end{eqnarray}

\begin{eqnarray}
&&\Pi ^{\mathrm{Dim10}}(M^{2},\alpha )=-\frac{m_{c}\langle \alpha
_{s}G^{2}/\pi \rangle g_{s}^{2}}{23328M^{4}\pi ^{2}(1-\alpha )^{3}}\exp %
\left[ -\frac{m_{c}^{2}}{M^{2}(1-\alpha )}\right] \left\{
m_{c}^{2}m_{s}\langle \overline{u}u\rangle ^{2}(1-\alpha )\right.  \notag \\
&&-2M^{2}m_{s}\langle \overline{u}u\rangle ^{2}(1-\alpha )+\alpha \lbrack
m_{c}^{3}-m_{c}M^{2}(1-\alpha )[\langle \overline{u}u\rangle ^{2}+\langle
\overline{s}s\rangle ^{2}]+\langle \overline{d}d\rangle ^{2}\left[
-2M^{2}m_{s}(1-\alpha )\right.  \notag \\
&&\left. \left. +m_{c}^{3}\alpha -m_{c}M^{2}\alpha (1-\alpha
)+m_{c}^{2}m_{s}(1-\alpha )\right] \right\} +\frac{m_{c}^{2}}{1728\pi
^{2}M^{4}}\exp \left[ -\frac{m_{c}^{2}}{M^{2}}\right] (m_{c}m_{s}+M^{2})
\notag \\
&&\times \left[ 9\langle \overline{u}g_{s}\sigma Gu\rangle \langle \overline{%
d}g_{s}\sigma Gd\rangle +8\pi ^{2}\langle \alpha _{s}G^{2}/\pi \rangle
\langle \overline{u}u\rangle \langle \overline{d}d\rangle \right].
\label{eq:A17}
\end{eqnarray}

In expressions above, $\Theta (z)$ is Unit Step function, and
\begin{equation}
\ L=s\alpha (1-\alpha )-m_{c}^{2}\alpha .
\end{equation}%
Calculation of the correlation function $\Pi (M^{2},s_{0})$ implies
integrations of the spectral density's components $\rho ^{\mathrm{pert.}}(s,\alpha )  $ and $\rho ^{\mathrm{%
Dim3(4,5,6,7,8)}}(s,\alpha )$ over $\alpha $ and $s$. These functions
contain $\sim 1/(1-\alpha )^{n}$ type factors, which at $\alpha =1$ may lead
to singularities and diverge integrals. These spectral densities, however, depend also on
the function $\Theta \lbrack s\alpha (1-\alpha )-m_{c}^{2}\alpha ]$ that
cuts off dangerous regions in $\alpha $ integration. The components of $\
\Pi (M^{2})$ are functions of the Feynman parameter $\alpha $, and have in
denominators the factors $(1-\alpha )^{n}$ as well. But the function $\exp %
\left[ -m_{c}^{2}/M^{2}(1-\alpha )\right] $ in these components effectively
regulates possible singularities rendering finite relevant integrals.

\end{widetext}

\end{document}